\documentclass[11pt,draftcls,onecolumn]{article}
\usepackage{multicol,caption}
\usepackage{multirow}
\usepackage[a4paper, total={7in, 9in}]{geometry}
\usepackage{graphicx}
\usepackage{adjustbox}
\usepackage{changepage}
\usepackage{enumitem}
\usepackage{listings}
\usepackage{amsmath}
\usepackage{color}
\usepackage{rotating}
\usepackage{pdfpages}
\usepackage{amsfonts}
\usepackage{tikz}
\tikzset{
	treenode/.style = {shape=rectangle, rounded corners,
		draw, align=center,
		top color=white, bottom color=blue!20},
	root/.style     = {treenode, font=\Large, bottom color=red!30},
	env/.style      = {treenode, font=\ttfamily\normalsize},
	dummy/.style    = {circle,draw}
}

\usepackage{listings}
\usepackage{courier}

\usepackage{natbib}
\usepackage{apalike}

\newenvironment{Figure}
{\par\medskip\noindent\minipage{\linewidth}}
{\endminipage\par\medskip}

\newcommand{\bc}{\begin{center}}
\newcommand{\ec}{\end{center}}

\begin{document}
	\title{Patterns of social mobility across social groups in India}
	\author{Vinay Reddy Venumuddala\\ PhD Student in Public Policy at IIM Bangalore.\\}
	\date{}
	\maketitle

\begin{quotation}
	\scriptsize
	``\textit{If all human beings are equally deserving of respect, and if respecting human beings means promoting their free agency, and if all human beings have capacities for agency that need development, and if society shapes the degree to which they can develop these capacities, and does so in particular by making resources available to them, and if, finally, society is a cooperative effort that we can shape and reshape if we wish, then we can remake the distribution of resources in our society so that it better helps all its members develop their capacities, and our obligation to respect other human beings entails that we should do so.}'' - \cite{fleischacker2009short}, Pg.123.
\end{quotation}

\section{Introduction}
Substantive research on inclusive growth emphasizes that, growth without equality of opportunity can propel individuals of a society into the detrimental consequences of rising inequalities, posing danger to its political and social stability \citep{rauniyar2010inclusive}. Empirically this alludes to the fact that, inequality of opportunity is detrimental to equity, a phenomenon whose  normative significance also emerges as a one of the recent culminations in the long drawn history of the evolution of distributive justice. The second formulation on distributive justice by John Rawls, proposes that if social and economic inequalities are to be arranged in a society, they must be to the greatest benefit of the least advantaged; and most importantly such inequalities, if needed to be attached to offices and positions, must necessarily entail them to be open to all under conditions of fair equality of opportunity (On Rawls's second formulation by, \cite{fleischacker2009short}, pg.114). An inquiry into the causes of rising (or persisting at the very least) socio-economic inequalities  in a society, therefore may not be easy if one ignores the importance of gauging equality/inequality of opportunity. In this context the well-known connotation of ``Great Gatsby Curve'', which shows that inequality is strongly associated with inter-generational earnings elasticity\footnote{Intergenerational earnings elasticity is measured as the elasticity between parental earnings and child's earnings \citep{corak2013income}. That is, it measures the extent to which a child's economic status is dependent on his/her parent.} across countries, is of particular importance. Intergenerational mobility measures which essentially indicate the opposite of status persistence across generations, are widely used as proxy for equality of opportunity \citep{chetty2014land,Asher}. The Great Gatsby Curve therefore signals to a strong negative relationship between economic inequality and equality of opportunity or intergenerational mobility, and thus points to the possibility that higher inequality together with poor intergenerational mobility may viciously lead to rising inequalities in the future \citep{corak2013income}. In the context of a larger inquiry into pathways through which inequality persists in a society portending an unsustainable growth, measuring social mobility\footnote{In the subsequent sections we use social mobility or intergenerational mobility interchangeably.} therefore is crucial.  \\

Furthermore, intergenerational status transmission which can happen through social and cultural means even beyond the less disguised forms of economic means, may be shaped by not just family circumstances alone, but also by the social structures and institutions which play a dominant role in the production of different forms of capital \citep{bourdieu2011forms}. In segmented societies like India, long-lasting institutions of caste and religion have not just unequalized the distribution of social and cultural capital, but also made some forms of them more scarce than others in the society. If we just consider economic capital alone, caste inequalities in India continue to exist even today and show up starkly in the most aspired occupations in the society. Its a truism that the varieties of occupations and professions into which individuals of different caste groups end up in, is much wider today than it was immediately post-independence. However it does not change the social reality that an overwhelming majority of the most preferred/sought after occupations in the society are held by individuals from upper castes, and those occupations which are considered most menial and despised, are held by individuals from lower castes (\cite{deshpande2003contemporary}, Ch.5). This reality which is shaped possibly by the nature of conversions among different forms of capital, is something that motivates us to look at intergenerational mobility along various dimensions (some of them being income, education and occupation). And most importantly, unless we compare the intergenerational mobility patterns across social groups, an overall outlook may simply be insufficient to comment on the true nature and extent of inequality in opportunity in the country.\\

For measuring intergenerational mobility, in this study, we stick mainly to two important dimensions which indicate the social status, namely occupation and education. We separately compute the mobility patterns across social groups along these dimensions, and do a compare and contrast. In the following sections, therefore we first characterize the measurement of social mobility in great detail, and subsequently look at the patterns of mobility across social groups, discuss our findings and conclude.\\

\section{Characterizing Social Mobility}
Social Mobility captures the extent to which socio-economic status of children, are independent of such status of their parents. In order to measure social mobility, most widely used indicators of socio-economic status are income, education and occupation. While social mobility measurement based on income is less contested, data availability in Indian context severely limits such approach \citep{Asher}. With education and occupation as variables associated with socio-economic status of individuals, measurement of mobility takes two dominant routes. Methods falling in the first route, are predominantly applied with occupations as discrete categories and use transition matrices in projecting the estimates of mobility \citep{motiram2012close,azam2015intergenerational,iversen2017rags}. These methods can also be applied with education as discrete categories. However a comparison of upward or downward mobility across social groups using transition matrix based methods is severely prone to subjective interpretation\footnote{It is subject to vary with variation in the scheme used for categorizing occupations for example.}. Methods in the second route measure mobility within regression framework, treating socio-economic status on a continuous scale whether its income, education or occupation. Within the second route, mobility measurement differs in interpretation when the said socio-economic status variable is coded directly or as an ordered rank in a particular time setting. If we consider income/education as the variables indicating socio-economic status, relative mobility measurement requires the information on the levels of these variables for both parent and child. Absolute mobility measurement however relies on the ranks of parent and child in their respective income/education distributions\footnote{For example, `Intergenerational income elasticity' which gives the measure of relative mobility can be measured through a regression of log of child's income/education on log of parent's income/education. Absolute mobility which is a rank based measure, can be computed, for example by finding the expected rank of children whose parents are at the 25th percentile of their income distribution. \citep{chetty2014land} }. In this study, we focus on absolute mobility measurement within the second route (motivated from the works of \cite{chetty2014land} and \cite{Asher}). Since mobility measurement through income is rather difficult to arrive at, owing to data limitations, we are left with treating either education or occupation on a continuous scale indicating socio-economic status.\\

 In this study we particularly focus on bringing occupations onto a continuous scale representing socio-economic status, in the process of computing absolute mobility measures. We also measure the mobility patterns across social groups using education measured on a continuous scale (Years of Schooling), which will help us to compare the results with existing studies. From our limited knowledge, ranking occupations onto a continuous scale for the purpose of absolute mobility measurement, is not attempted before in the Indian context, and hence the inferences may be new and perhaps interesting. Before venturing into the methodological details, a short description of the datasets used is given below.\\
 
 
   We use two major datasets for the purpose of our study, first one is Indian Human Development Survey - II (IHDS-2) which has information on education and occupation of not just parent-child pairs within the household, but also of non co-resident\footnote{Co-resident implies both parent and child are residing in the same observed household, while non-coresident hosueholds may have either parent or child living outside the hosuehold, but their information is captured in the survey.} father of the head of the household. Since such information pertaining to non co-resident mother has not been captured by the survey, we restrict ourselves to only father-son pairs in reporting the mobility patterns. Second dataset that we use is NSSO consumption expenditure survey, 43rd round (1988-90). In order to analyse differences in mobility through time, we divide father-son pairs from IHDS-2, on the basis of decadal birth cohorts of the working sons. For computing the socio-economic status ranking of occupations corresponding to each birth cohort, which we dwell upon in detail in subsequent sections, data from both the surveys are utilized. We use IHDS2 to arrive at the occupational ranking for most recent birth cohorts, and NSSO 43rd round on consumption expenditure for the older cohorts. Following table (Table-\ref{Cohorts}) indicates the birth cohorts and the corresponding dataset used for arriving at occupational ranking pertinent to each birth cohort.\\

\begin{table}[!ht]
	\centering
	\begin{tabular}{|l|c|c|c|c|c|c|}
		\hline
		Birth-Cohorts & 1926-35 & 1936-45 & 1946-55 & 1956-65 & 1966-75 & 1976-85 \\ 
		\hline
		Age (in 1988) &53-62  &43-52  & 33-42 & 23-32&- & - \\ 
		\hline
		Age (in 2011) & - & - & - & - & 36-45 & 26-35 \\ 
		\hline
		SEI computed from & NSS43 & NSS43 & NSS43 & NSS43 & IHDS2 &IHDS2\\ 
		\hline
	\end{tabular}
	\caption{\label{Cohorts} Cohorts\\ \scriptsize (SEI represents socio economic index scores for occupations, which we discuss in detail in the subsequent sections)}
\end{table}

In the following section we first evaluate the significance of using occupation as the dimension representing socio-economic status, and later illustrate one important method for obtaining socio-economic status scores on a continuous scale, for the occupations\footnote{Both IHDS-2 and NSSO 43rd Round use NCO-1968 classification of occupations. However since IHDS-2 uses 2-digit and NSS uses 3-digit codes, we re-map NSS codes to 2-digit in order maintain compatibility across datasets.} we have.

\section{Occupations on a continuous scale}

While intergenerational mobility measured in education is comparable across countries, education in itself may not reflect the true socio economic status of an individual. For the same level of education it is commonly observed that, across individuals the level of income and/or the consequent socio-economic status varies rather starkly. Income mobility measurement also has its share of limitations. Assumptions that income or economic capital can characterise socio-economic status can be inappropriate in the settings where cultural and social capital are strongly associated with such status. In fact most historical studies on mobility and stratification use occupation of an individual as a good indicator of social position instead of education or income. Even though such usage is motivated by the availability of data on individual occupations in historically recorded censuses, it also recognizes the problems with immanent variability, low and high respectively  observed along the other dimensions i.e., education and income, both in the past and in the present times \citep{van2010historical}. \\

However in recent times measuring mobility using occupation of individuals has not been appreciated well enough relative to other dimensions like income or education. One main reason for it is the difficulty in parsimoniously characterizing mobility from intergenerational occupational transitions. While transition matrix based methods provide a means to do that, their interpretation is more descriptive and therefore makes it less attractive for making mobility comparisons across social groups and much more difficult to carry out comparisons across countries. Nevertheless some of the recent works on occupational mobility in India by \cite{motiram2012close}, \cite{azam2015intergenerational} and \cite{iversen2017rags} have been quite influential, particularly in describing the mobility patterns across social groups, despite such limitations. \\

In this paper, our objective is to put occupations on a continuous scale, so as to bring this dimension into the field of most recent methods on mobility, particularly suited for comparative studies (across social groups and even countries). Recent methods, particularly those motivated by \cite{chetty2014land}, 
allow for absolute mobility measurement with information on income or education of children and their respective parents, irrespective of the support of their corresponding distributions. Percentile rank of sons within their status\footnote{Status may refer to income/education/occupation here.} distribution are regressed on the percentile rank of their respective fathers within the latter's distribution. From this regression, an estimate of son's rank given a particular rank of father allows us to infer about absolute mobility. In a perfectly mobile society, expected rank of a son, for his father in any percentile rank in the national status distribution is 50. For other societies, \textbf{a.} The conditional expectation of son's percentile rank given father is in  25th percentile of his distribution, and \textbf{b.} The conditional expectation of son's percentile rank given father is in 75th percentile of his distribution, both of them can be compared by their distance to the perfect scenario (i.e 50). While the former represents absolute upward mobility, latter represents absolute downward mobility. This scheme is compatible for mobility comparisons across countries or social groups  \citep{chetty2014land,Asher}. Mathematical representation of this scheme is given in Section-\ref{sec:MobilityGroups}.\\

 In order to reap advantages from the above mentioned scheme, we first map occupations onto a continuous scale by using socio-economic-index\footnote{\cite{ganzeboom2003three} summarizes three internationally standardized measures for comparative research on occupational status. First two, Prestige based measures and socio-economic indices, rate occupations on a continuous scale, while the last one EGP-class (Erikson - Goldthorpe - Portocarero) places occupations under nominal class categories. Within the continuous class of measures representing occupational status, empirical evidence favours socio-economic status indices over prestige based ones, in terms of occupations driving the status attainment process (\cite{ganzeboom2003three} and \cite{ganzeboom1992standard}).} (SEI) scores for occupations, as proposed by \cite{ganzeboom1992standard}. Like education, but more close to reflecting socio-economic status, SEI measures the attributes of occupations that convert a person's education into income \citep{ganzeboom1992standard,ganzeboom2003three}. In order to account for the changing socio-economic status of occupations with time, we  divide individuals (irrespective of whether they are fathers/sons) into birth-cohorts, and compute SEI scores separately by using attributes of income, education and occupation of the individuals falling within each birth cohort. Given the nature of our datasets\footnote{Individual incomes are not reported in either of the datasets we use (NSS43, and IHDS2). We therefore rely on the available household level information.}, we focus on household heads within each birth cohort as our individual units; their education, principal occupation, and corresponding monthly household income are used for the computation of cohort-specific SEI scores. Further in order to make the SEI scores across cohorts compatible, we rescale them in the range $(1,80)$.\footnote{ Rescaling is also motivated from a similar cross-country SEI score comparison conducted in \cite{ganzeboom2003three}. 80 is the total number of distinct occupation codes pertinent to the father-son pairs used for mobility comparisons in our study using IHDS data. (Total father-son pairs across all cohorts in our sample, is above 30K. See Table - \ref{CrossTableFS})} Following steps shown below roughly sketches the overall procedure we adopt for bringing occupations into the fold of recent methods in mobility measurement.
  
 \begin{enumerate}
	\item Classify individuals (household heads, irrespective of whether father or son) on the basis of their birth cohort ($C_t$).  
	\item Compute SEI scores for occupations corresponding to individuals (heads) in $C_t$, using their attributes for $I$ (Income), $E$ (Education), and $O$ (occupation), as discussed previously. Details of the method are given in Section-\ref{ComputeSEI}.
	\item Rescale cohort specific SEI scores to range between (1,80). We denote these rescaled scores by $R_{it}$, where $i$ indicates occupation and $t$ indicates birth cohort). 
	\item For every individual in the father-son pairs obtained from IHDS-2 dataset (which have information on education and occupation separately for fathers and sons), we assign $R_{it}$ on the basis of their corresponding birth cohort and occupation.\footnote{As a part of our robustness, we also compute $R_i$ = $(1/n)*\sum_{t=1}^{n(\#Cohorts)} R_{it}$, which is a cohort-averaged  status score of each occupation, to compare the mobility patterns it generates in comparison to our main scheme.} 
	\item Like income or education, we therefore have rescaled SEI scores corresponding to each father and son in the data of our total father-son pairs. These scores help us to lay out the status distributions of sons and their respective fathers separately, and therefore allows us to order sons and fathers on the basis of percentile ranks within their respective distributions.\footnote{It is important to note that these percentile ranks are computed for sons belonging to all the social groups in a birth cohort and therefore allows us to make mobility comparison across groups in the lines of \cite{chetty2014land} and \cite{Asher}.} 
	\item Using the percentile ranks for each birth cohort, thus measured on a socio-economic status scale found on individual occupations (similar to income or education), we compute absolute upward and downward mobility measures across social groups similar to \cite{chetty2014land}.
\end{enumerate}

\subsection{Computing SEI scores} \label{ComputeSEI}
In order to compute socio-economic-index (SEI) scores for occupations, we adopt the method suggested by \cite{ganzeboom1992standard}. According to them, SEI measures the attributes of occupations that, \textit{converts a person's main resource (education) into a person's main reward (income)}. It is defined as the, \textit{intervening variable between education and income that maximizes the indirect effect of education on income and minimizes the direct effect} \citep{ganzeboom1992standard}.\\

Figure-\ref{fig:SAM} shows the basic status attainment model with occupation as an intervening variable as proposed by \cite{ganzeboom1992standard}. Algorithm\footnote{Alternating least alternating least square algorithm to minimize direct effect in a path model, proposed by \cite{de1988multivariate} as cited in \cite{ganzeboom1992standard}} to arrive at SEI scores is given in table-\ref{SAMAlgo}. We run this algorithm separately for heads of the households belonging to each birth cohort\footnote{Using their corresponding data from NSSO and IHDS. As shown in the table-\ref{Cohorts} previously, we use NSS data for finding SEI scores for occupations across the first four birth cohorts, and IHDS2 for the last two cohorts.}, in order to compute the cohort wise SEI scores for each occupation.  For this purpose as mentioned previously, we use  information on principal occupation of the household ($O_i$), Monthly household consumption expenditure as proxy for income (`INC'), and Age and Education\footnote{NSS43 presents education of individuals in discrete levels. We recode these discrete levels of education on a continuous scale using InSCED mapping (Indian Standard Classification on Eduacation).
	\scriptsize Recoding Scheme: ``\textit{NSS Code (Recoded Years of Schooling):} 0-illiterate (0), 1-literate without formal schooling (1), 2-literate but below primary (3), 3-primary (8), 4-middle (11), 5- secondary (13), 6-9-Graduates (17)''} of the head of the household (`AGE' and `EDU').


\begin{Figure}
	\captionsetup{font=scriptsize}
	\begin{center}
		\includegraphics[width=2.0in,angle=-90]{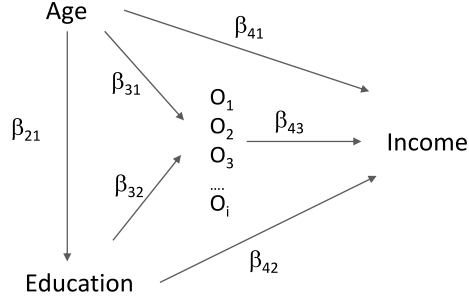}
		\captionof{figure}{Basic Status Attainment Model (Recreated from \cite{ganzeboom1992standard})} 
		\label{fig:SAM}
	\end{center}	
\end{Figure}

\begin{table}[ht]
	\centering
	\begin{tabular}{|l|p{15cm}|}
		\hline
		\textbf{S.No} & \textbf{Description} \\  
		\hline
		Step 1: & Initialize the education and income weights at any reasonable starting point (e.g., 0.5 and 0.5) and construct a starting point.\\
		\hline
		Step 2: & 1. Regress INC on AGE and SEI ($\beta_{41},\beta_{43}$) (*) \newline 2. Regress SEI on EDU and AGE ($\beta_{32},\beta_{31}$) \newline 3. Regress EDU on AGE ($\beta_{21}$)\\
		\hline
		Step 3: & 1. Compute $SEI' = \beta_{43}(INC-\beta_{41}AGE) + \beta_{32}EDU+\beta_{31}AGE$ \newline 2. Standardize $SEI'$ \newline 3. Compute scores as means of $SEI'$ for $O_1,O_2,..,O_i$ \newline 4. Compute $SEI''$ using the new scaling\\
		\hline
		Step4: & 1. Regress INC on AGE, EDU and SEI'' ($\beta_{42}$ is the coefficient on EDU)\newline 2. If minimum on $\beta_{42}$, step out \newline 3. Go back to Step 2 and substitute SEI'' for SEI\\
		\hline
	\end{tabular}
	\caption{\label{SAMAlgo} Algorithm for estimating an optimally scaled occupation variable, SEI, for the model in Figure-\ref{fig:SAM} (Recreated from \cite{ganzeboom1992standard})\\ \footnotesize Note:- AGE: age; INC: income; EDU: education; SEI, SEI',SEI''; estimated socio-economic index of occupational status. All variables need to be standardized with mean 0 and standard deviation 1.\\ (*) EDU is not included in this regression.}
\end{table}

The crucial coefficients that relate the resultant occupation status to education ($\beta_{32}$) and occupation status to income ($\beta_{43}$) according to Step-3 of the procedure, are given below (Table-\ref{Coeff}) for each birth cohort. From the table it is observed that the SEI scores of occupations across cohorts, are greatly determined by the income that they fetch, in comparison to the education level needed. Another important observation is that this strength of income in indicating occupational status score is increasing with time and is considerably high for the recent birth cohort.\\

\begin{table}[!ht]
	\centering
	\begin{tabular}{|p{3cm}|p{2cm}|p{2cm}|}
		\hline
		\textbf{Birth-Cohort} & $\beta_{43}$ & $\beta_{32}$ \\ 
		\hline
		1926-35 & 0.80 & 0.31 \\
		\hline
     	 1936-45 &0.78&0.37  \\ 
		\hline
		1945-56 &0.80  & 0.37 \\
		\hline
		 1956-65 &  0.79&0.31  \\ 
		 \hline
		1966-75 &0.80&0.27\\
		\hline
		1976-85 &0.86&0.20\\
		\hline
	\end{tabular}
	\caption{\label{Coeff} Strength of Income and Education in determining SEI score \\ \scriptsize (Refer Step-3 in Table-\ref{SAMAlgo})}
\end{table}

For a simplistic representation of obtained SEI scores across birth-cohorts, owing to paucity of space, we average the rescaled SEI scores obtained for 2-digit occupation classes on to corresponding 1-digit occupation codes. Table-\ref{tab:1DigSEI}, shows the average SEI scores, average Income and education ranks mapped from 2-digit to 1-digit occupation codes. Detailed scores for 2-digit occupation codes averaged across cohorts are shown in Table-\ref{tab:FullSummary} of Appendix (Section-\ref{sec:Appendix}). Change in SEI scores, income and education ranks for broad 1-digit occupation classes are clearly visible across birth cohorts in Table-\ref{tab:1DigSEI}. \\

\begin{table}[ht]
	\centering
	\begin{adjustbox}{width=1\textwidth}
		\begin{tabular}{ccccccccccccccccccc}
			\hline
			Code & 26R & 36R & 46R & 56R & 66R & 76R & $\overline{I26}$ & $\overline{I36}$ & $\overline{I46}$ & $\overline{I56}$ & $\overline{I66}$ & $\overline{I76}$ & $\overline{E26}$ & $\overline{E36}$ & $\overline{E46}$ & $\overline{E56}$ & $\overline{E66}$ & $\overline{E76}$ \\ 
			\hline
			0 & 62 & 64 & 61 & 61 & 61 & 60 & 9 & 9 & 9 & 9 & 10 & 10 & 9 & 10 & 10 & 10 & 9 & 10 \\ 
			1 & 55 & 50 & 57 & 51 & 49 & 46 & 8 & 7 & 8 & 7 & 7 & 7 & 8 & 7 & 9 & 7 & 10 & 9 \\ 
			2 & 64 & 65 & 64 & 66 & 58 & 44 & 10 & 10 & 10 & 10 & 9 & 8 & 10 & 9 & 8 & 8 & 7 & 7 \\ 
			3 & 51 & 52 & 52 & 53 & 50 & 47 & 7 & 8 & 7 & 8 & 8 & 9 & 7 & 8 & 7 & 9 & 8 & 8 \\ 
			4 & 42 & 43 & 42 & 43 & 41 & 38 & 6 & 6 & 6 & 6 & 6 & 6 & 6 & 6 & 6 & 6 & 6 & 6 \\ 
			5 & 22 & 19 & 22 & 26 & 29 & 25 & 1 & 1 & 2 & 3 & 4 & 2 & 2 & 2 & 2 & 2 & 3 & 5 \\ 
			6 & 21 & 20 & 21 & 20 & 17 & 19 & 2 & 2 & 1 & 1 & 1 & 1 & 1 & 1 & 1 & 1 & 1 & 1 \\ 
			7 & 29 & 28 & 28 & 27 & 26 & 27 & 5 & 4 & 4 & 4 & 2 & 4 & 3 & 3 & 4 & 3 & 2 & 2 \\ 
			8 & 28 & 32 & 30 & 31 & 31 & 28 & 3 & 5 & 5 & 5 & 5 & 5 & 4 & 5 & 5 & 5 & 5 & 4 \\ 
			9 & 29 & 26 & 25 & 26 & 29 & 25 & 4 & 3 & 3 & 2 & 3 & 3 & 5 & 4 & 3 & 4 & 4 & 3 \\ 
			\hline
		\end{tabular}
	\end{adjustbox}
	\caption{Average rescaled SEI scores shown for 1-digit occupation codes\\
		\scriptsize 
		\textit{\textbf{Codes}:- 0,1 - Professional, Technical and Related Workers; 2 - Administrative Executive and Managerial Workers; 3 - Clerical and Related Workers; 4 - Sales Workers; 5 - Service Workers; 6 - Farmers, agriculture labour, fisherman, hunters and related workers; 7,8,9 - Production and Related Workers, Transport Equipment Operators, Labourers;}\\
		\textit{\textbf{Note}:- yyR represents the average of rescaled SEI scores, averaged over all the 2-digit coded occupations within the corresponding 1-digit codes; $\overline{Iyy}$,$\overline{Eyy}$, represents the ranks of, average income and average education computed over all the 2-digit coded occupations within the corresponding 1-digit codes; yy indicates the son birth cohort (E.g. `1926-35' written as `26').}}
	\label{tab:1DigSEI}
\end{table}

In order to assess the extent of changes in occupational SEI scores (for 2-digit codes) across cohorts, we compute the correlation of SEI scores between consecutive cohorts, results of which are shown in Table-\ref{Corr}. One may roughly infer a little less flux in occupational status rankings for birth cohorts in 1926-1965, and a greater flux after (more on the basis of Kendall and Spearman coefficients which are rank based correlation measures). The political environment during 1950s-1980s, years which have seen an almost stagnant growth rate (of about 3.5\%) can be one possible reason which may explain this observation. Nevertheless, even if we ignore such changes in status of occupations over time, we still observe that the mobility patterns across social groups more or less display a consistent comparative picture, details of which we discuss in the next section (Section-\ref{OccRobust}).

\begin{table}[ht]
	\centering
	\begin{tabular}{|r|l|c|c|c|}
		\hline
		& Birth Cohort & Corr($S_t,S_{t-1}$) (Pearson) & Corr($S_t,S_{t-1}$) (Spearman) & Corr($S_t,S_{t-1}$) (Kendall) \\  
		\hline
		1 & 1926-35 & 0.00 & 0.00 & 0.00 \\ 
		2 & 1936-45 & 0.94 & 0.93 & 0.79 \\ 
		3 & 1946-55 & 0.96 & 0.95 & 0.84 \\ 
		4 & 1956-65 & 0.96 & 0.97 & 0.86 \\ 
		5 & 1966-75 & 0.88 & 0.90 & 0.73 \\ 
		6 & 1976-85 & 0.82 & 0.85 & 0.68 \\ 
		\hline
	\end{tabular}
\caption{\label{Corr} Correlation of Socio-Economic Status Rankings of occupations across birth cohorts\\ \footnotesize Note: $S_t$ is the vector of socio economic status ranking of occupations corresponding to a birth cohort and $S_{t-1}$ indicates its previous cohort.}
\end{table}

\section{Mobility patterns across social groups} \label{sec:MobilityGroups}
We observe absolute upward mobility as expected son's rank given a father at the 25th percentile in the parent rank distribution; and similarly, absolute downward mobility as the expected son's rank given a father at the 75th percentile in parent rank distribution. These definitions are motivated by the respective definitions from \cite{chetty2014land}. Following is the population regression equation which helps us to arrive at the mobility measures.
\begin{align*}
Y = \beta_0 + \beta_1X + \epsilon
\end{align*}
where $Y$ denotes the percentile rank of son, and $X$ denotes the percentile rank of his father. Estimated value of $Y$ at a particular value of $X=i$, denoted by $p_i = E(Y|X=i)$, indicates the absolute mobility measure at percentile $i$. In this study, we measure both the conventional upward and downward absolute mobility measures $p_{25}$ and $p_{75}$ following the work of \cite{chetty2014land}. Further, mobility patterns are separately computed with occupation (rescaled SEI scores) and education as the dimension indicating socio-economic status. \\

Mobility estimates and confidence bounds are arrived at using boot-strapping method. For each son birth cohort and dimension of interest, following are the steps that allow us to arrive at confidence bounds and approximate point estimates for mobility.
\begin{enumerate}
	\item If there are (say) `$N_t$' observations\footnote{Each observation has the information about education and occupation of son and his father. For each occupation, depending on the individual's birth cohort, we assign rescaled SEI scores as described previously.} within a son birth cohort, we obtain $N_t$ re-samples from this data with replacement.
	\item For each re-sample, we compute the point estimates of $p_{25}$ (= $E(Y|X=25)$) and $p_{75}$ (= $E(Y|X=75)$), for upward and downward mobility respectively.
	\item We arrive at the sampling distribution of $p_{25}$ and $p_{75}$ from the re-sampled estimates. Our point estimates for upward and downward mobility respectively are the corresponding mean of re-sampled estimates of $p_{25}$ and $p_{75}$. 
	\item Upper and Lower confidence limits are obtained by assuming normal distribution structure for these sampling distributions.
\end{enumerate}

We find that the patterns of upward and downward mobility in education estimated by this method are very close to the ones presented in \cite{Asher}, as we show in Section-\ref{subsec:EM}. In their approach, absolute upward mobility is defined as $E(Y|X\in(0,50))$, i.e., expected son's rank for parents in the bottom half of their status distribution; and absolute downward mobility is defined as $E(Y|X\in(50,100))$, i.e., expected son's rank for parents in the top half of their status distribution. Nevertheless, the trends generated by these interval mobility estimates are by definition comparable to the ones generated by using $p_{25}$ and $p_{75}$, only with varying accuracy as discussed in \cite{Asher}. 

\subsection{Occupation Mobility Patterns} \label{subsec:OM}

Following tables - \ref{UM} and \ref{DM}, show the occupation upward and downward mobility point estimates across social groups for each son birth-cohort, computed using rescaled SEI scores as discussed in previous sections. Subsequent figures - \ref{fig:UMO} and \ref{fig:DMO} depict the same mobility patterns graphically, along with confidence intervals (indicated by the vertical segments within the respective plots). Observed patterns indicate that, for most recent cohorts whose bounds are relatively sharper than older cohorts\footnote{This finding in the context of education is elaborately explained in the work of \cite{Asher}. To briefly summarize, when majority of the fathers corresponding to son's in the older cohorts, belong predominantly to one or two education/occupation level(s), the corresponding bounds get coarser in such situation.}, both upward and downward mobility for SCs and STs are much lower than that of any other social group. These results hold even after robustness checks as we will see below in Section-\ref{OccRobust}. \\

\begin{table}[!ht]
	\centering
	\begin{tabular}{|l|llll|}
		\hline
		& 1946-55 & 1956-65 & 1966-75 & 1976-85 \\ 
		\hline
		Brahmin & 55.78, (n=169) & 52.12, (n=446) & 52.37, (n=449) & 52.74, (n=358) \\ 
		FC & 43.32, (n=580) & 50.36, (n=1558) & 46.09, (n=1558) & 45.85, (n=1328) \\ 
		OBC & 39.35, (n=1099) & 41.66, (n=2898) & 39.72, (n=3468) & 42.32, (n=3259) \\ 
		Dalits & 35.33, (n=588) & 34.73, (n=1701) & 36.81, (n=2179) & 36.87, (n=2259) \\ 
		Adivasis & 37.95, (n=288) & 35.03, (n=767) & 31.23, (n=873) & 34.53, (n=891) \\ 
		Muslims & 39.56, (n=307) & 40.54, (n=909) & 45.46, (n=1075) & 46.17, (n=1124) \\ 
		Others & 47.98, (n=93) & 51.76, (n=266) & 44.34, (n=262) & 53.77, (n=182) \\ 
		\hline
	\end{tabular}
	\caption{\label{UM} Upward Occupation Mobility (Mean Estimates)\\
		\footnotesize Note: The value of `n' in the parentheses indicates the number of father-son pairs used in the computation of corresponding mobility statistic. However, please note that appropriate household weights are used while computing the percentile ranks for each of the father and son occupations while coming up with these estimates. Columns correspond to each birth cohort.}
\end{table}

\begin{table}[!ht]
	\centering
	\begin{tabular}{|l|llll|}
		\hline
		& 1946-55 & 1956-65 & 1966-75 & 1976-85 \\ 
		\hline
		Brahmin & 69.96, (n=169) & 71.62, (n=446) & 70.97, (n=449) & 69.32, (n=358) \\ 
		FC & 66.4, (n=580) & 66.84, (n=1558) & 66.58, (n=1558) & 64.16, (n=1328) \\ 
		OBC & 61.59, (n=1099) & 59.59, (n=2898) & 59.84, (n=3468) & 59.65, (n=3259) \\ 
		Dalits & 57.3, (n=588) & 56.48, (n=1701) & 55.71, (n=2179) & 54.86, (n=2259) \\ 
		Adivasis & 59.71, (n=288) & 57.94, (n=767) & 55.63, (n=873) & 50.47, (n=891) \\ 
		Muslims & 56.8, (n=307) & 58.93, (n=909) & 64.27, (n=1075) & 64.94, (n=1124) \\ 
		Others & 64.48, (n=93) & 64.79, (n=266) & 64.21, (n=262) & 67.98, (n=182) \\ 
		\hline
	\end{tabular}
	\caption{\label{DM} Downward Occupation Mobility (Mean Estimates)\\
		\footnotesize Note given in Table-\ref{UM} applies here.}
\end{table}

\begin{table}[!ht]
	\begin{minipage}{.5\columnwidth}
		\begin{Figure}
			\captionsetup{font=scriptsize}
			\begin{center}
				\includegraphics[width=3.5in]{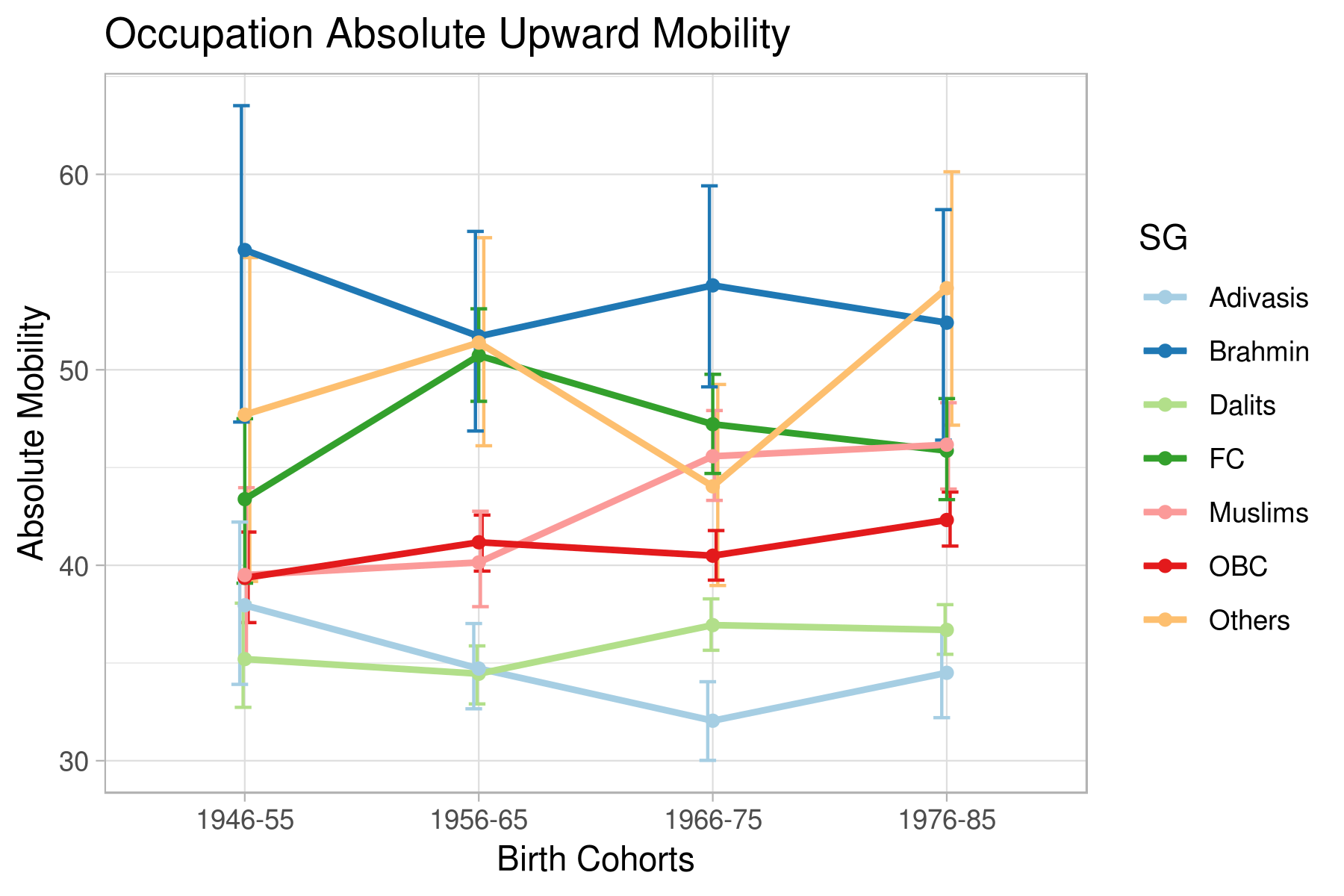}
				\captionof{figure}{Upward Mobility Estimates (Occupation)} 
				\label{fig:UMO}
			\end{center}	
		\end{Figure}
	\end{minipage}
	\begin{minipage}{.5\columnwidth}
		\begin{Figure}
			\captionsetup{font=scriptsize}
			\begin{center}
				\includegraphics[width=3.5in]{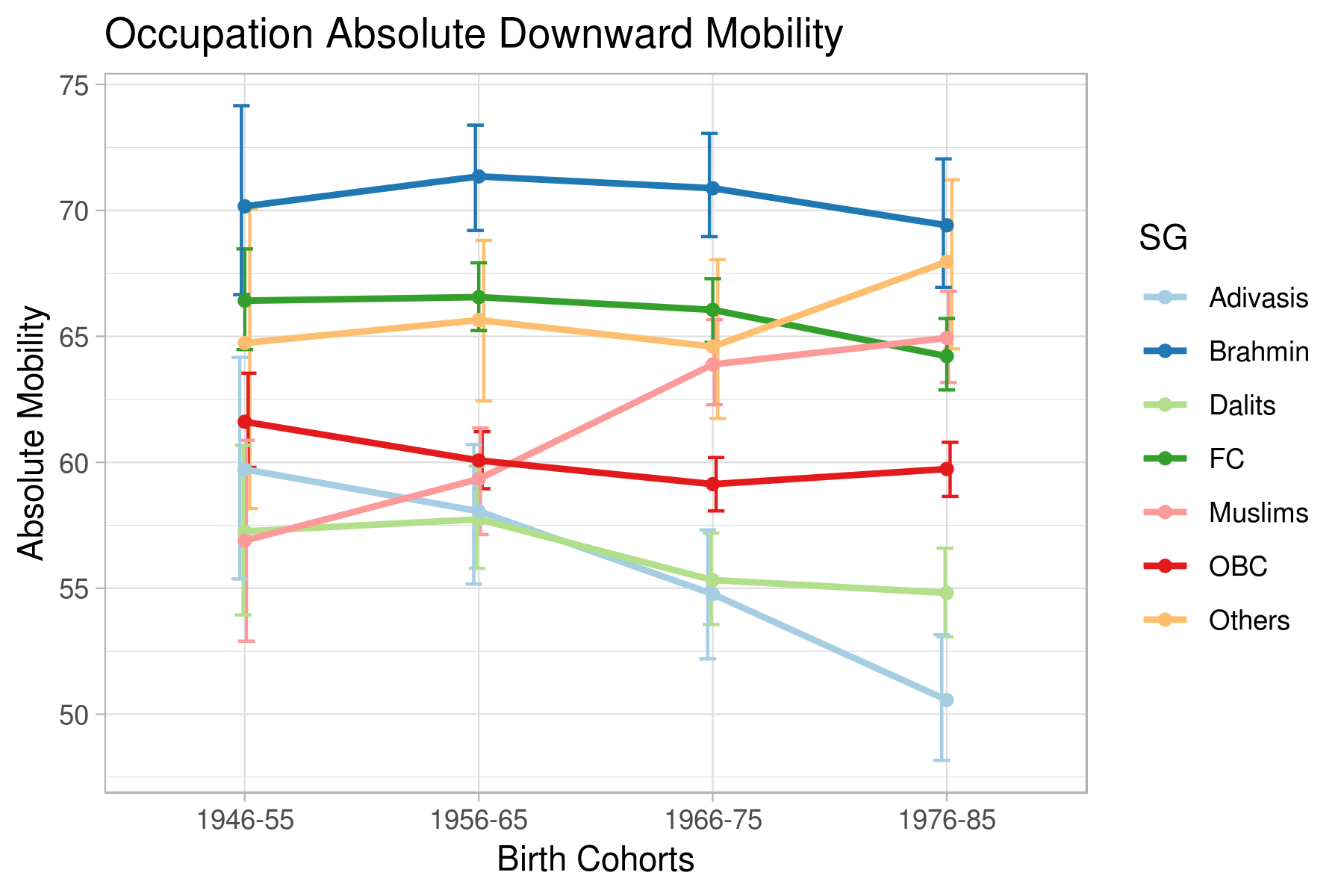}
				\captionof{figure}{Downward Mobility Estimates (Occupation)} 
				\label{fig:DMO}
			\end{center}	
		\end{Figure}
	\end{minipage}
\end{table}

\subsubsection{Robustness Checks} \label{OccRobust}
In order to check for robustness of our methodology, we observe occupation mobility patterns across social groups by deviating on following factors independently
\begin{enumerate}
	\item First, instead of using rescaled SEI scores separately for each birth cohort, we use average of these rescaled SEI scores across cohorts, to arrive at cohort independent SEI scores for all occupations (See Appendix-\ref{tab:FullSummary}, for the scores). Using these scores as reflecting the socio-economic status of individuals belonging to corresponding occupations, we arrive at mobility patterns.
	\item  Second, we modify the birth cohorts by shifting each cohort by 4 years. Tables - \ref{CrossTableFS} and \ref{CrossTableFS_Shift} indicate the number of father-son pairs in the respective birth cohorts for our original scheme, and new scheme respectively.
\end{enumerate}

In so far as the above robustness checks are concerned, we find that our occupation mobility trends show similar patterns across social groups except for slight differences in the recent cohorts. Nevertheless these differences are more or less visible only in the recent birth cohorts, and therefore are indicative of a continuing trend, inclusive of the most recent workforce.\footnote{Recent son birth cohort in original scheme is 1976-85, and in the new scheme it is 1980-89. Therefore the continuing trends pointed by this alternate scheme, are inclusive of individuals belonging to smaller cohort 1986-89.}

\begin{table}[ht]
	\centering
	\begin{tabular}{|c|c|c|c|c|c|}
		\hline
		\textbf{SC/FC}&  1926-35 &    1936-45 &     1946-55 &     1956-65 & Row Total  \\  
		\hline
		1946-55 &     3126 &        0 &         0   &         0 &      3126  \\
		\hline
		1956-65 &     3914 &      4634 &         0 &         0 &     8548  \\
		\hline
		1966-75 &         0 &      4957 &      4908 &         0 &     9865  \\
		\hline
		1976-85 &         0 &         0 &      4577 &      4827 &     9404  \\
		\hline
		Column Total &      7040 &      9591 &      9485 &      4827 &    30943  \\
		\hline
	\end{tabular}
	\caption{\label{CrossTableFS} Cross Tabulation of father-son birth cohorts in the sample.\\
		\footnotesize Note: Row Names indicate birth cohorts corresponding to sons and column names indicate corresponding birth cohorts of their fathers}
\end{table}

\begin{table}[ht]
	\centering
	\begin{tabular}{|c|c|c|c|c|c|}
		\hline
		\textbf{SC/FC}&  1930-39 &   1940-49 &   1950-59 &   1960-69 & Row Total   \\  
		\hline
		1950-59 &       3971 &         0 &        0 &         0 &    3971\\
		\hline
		1960-69 &     4329 &      5216 &       0 &      0 &      9545\\
		\hline
		1970-79 &          0 &      4552 &      4849 &        0 &      9401  \\
		\hline
		1980-89 &         0 &        0 &    4437 &  4674 &     9111 \\ 
		\hline
		Column Total &      8300 &     9768 &      9286 &     4674 &     32028 \\
		\hline
	\end{tabular}
	\caption{\label{CrossTableFS_Shift} Cross Tabulation of a new set of father-son birth cohorts in the sample.\\
		\footnotesize Note: Row Names indicate birth cohorts corresponding to sons and column names indicate corresponding birth cohorts of their fathers}
\end{table}

Figures - \ref{fig:UMOR} and \ref{fig:DMOR}, depict occupational mobility trends across social groups with constant SEI scores for all occupations across cohorts, which forms our first robustness check. Results of our second robustness check, are depicted in figures - \ref{fig:UMO_Shift} and \ref{fig:DMO_Shift} in Appendix (Section-\ref{sec:Appendix}). Together, these results more or less show similar comparative trends in occupational mobility across social groups like our original scheme, and therefore strengthen our confidence in the observed mobility patterns. Our results consistently show that, for the recent birth cohorts, lowest occupational mobility prospects are observed for STs, followed by SCs, then OBCs and Muslims. Please note that, owing to sample size concerns (See Tables-\ref{UM},\ref{DM}), mobility trends of SCs, STs, Muslims, OBCs,and FCs, in the recent birth cohorts are the most reliable. Trends observed for others need to be interpreted with caution. Nevertheless, we focus on the mobility trends mainly of SCs, STs, Muslims and OBCs in our study, and based on the conspicuously visible mobility trends (along both dimensions, occupation and education) for the other groups, we put them ahead of these four, on all measures.

\begin{table}[!ht]
	\begin{minipage}{.5\columnwidth}
		\begin{Figure}
			\captionsetup{font=scriptsize}
			\begin{center}
				\includegraphics[width=3.5in]{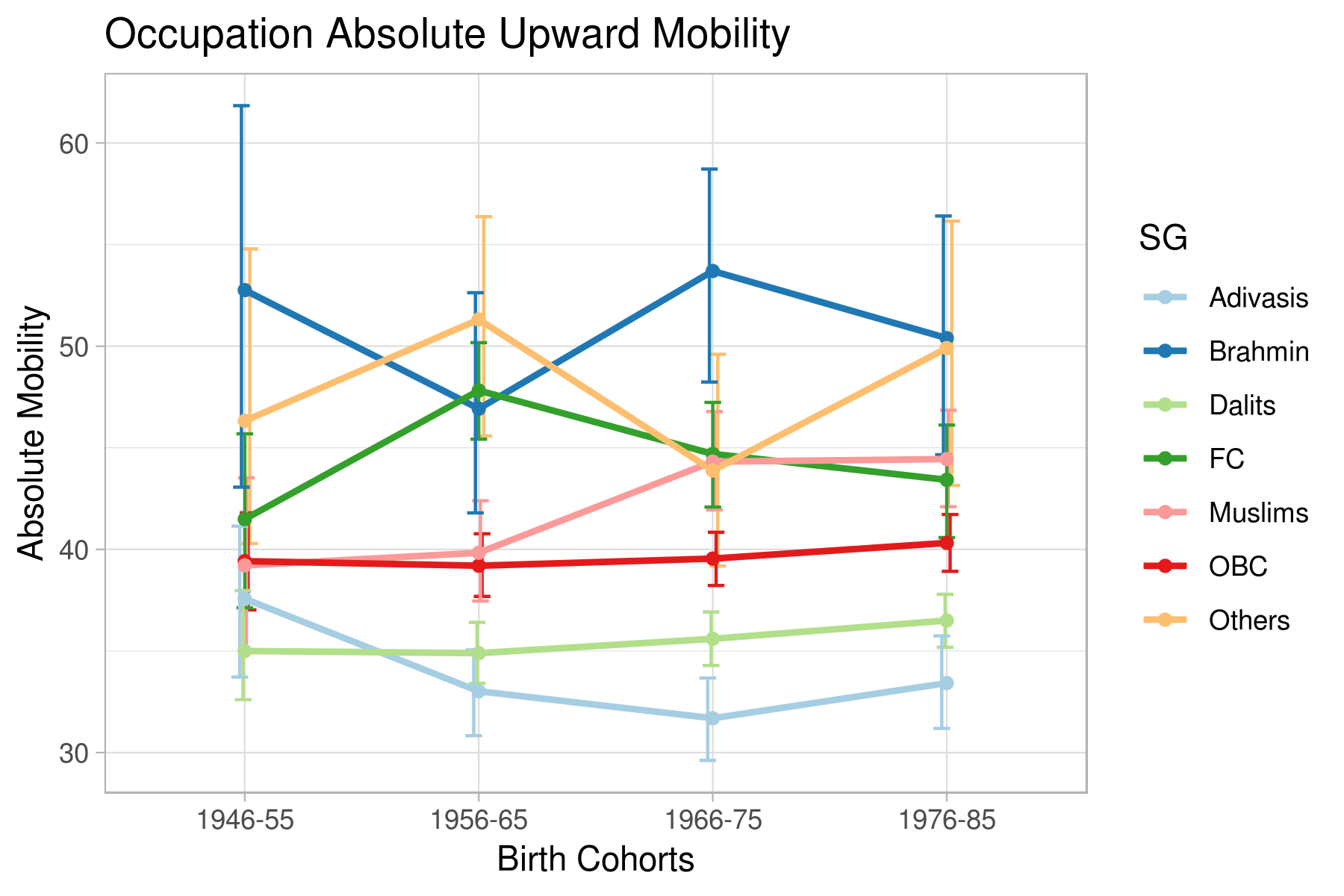}
				\captionof{figure}{Upward Mobility Estimates (Occupations with constant SEI ranks across cohorts)} 
				\label{fig:UMOR}
			\end{center}	
		\end{Figure}
	\end{minipage}
	\begin{minipage}{.5\columnwidth}
		\begin{Figure}
			\captionsetup{font=scriptsize}
			\begin{center}
				\includegraphics[width=3.5in]{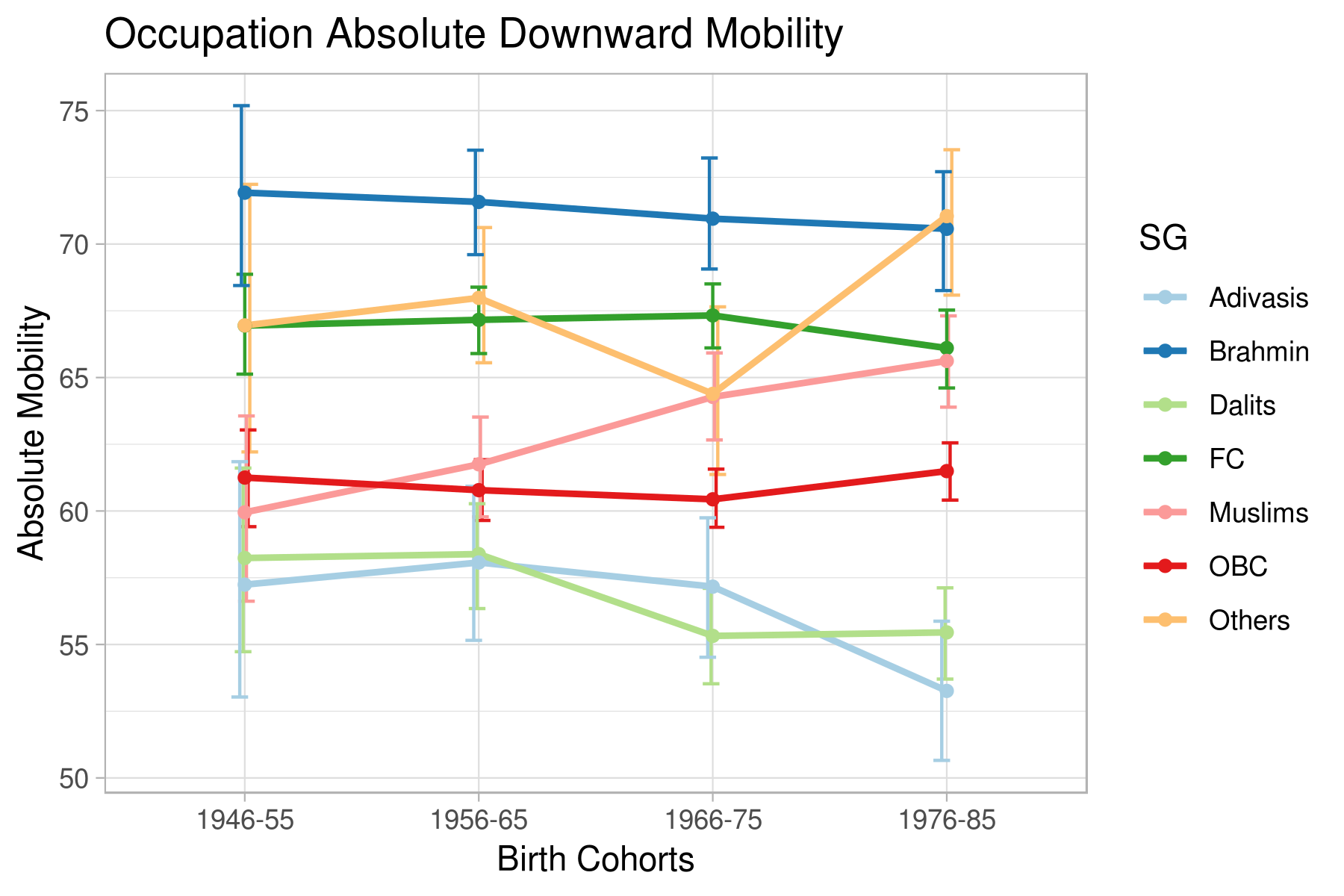}
				\captionof{figure}{Downward Mobility Estimates (Occupations with constant SEI ranks across cohorts)} 
				\label{fig:DMOR}
			\end{center}	
		\end{Figure}
	\end{minipage}
\end{table}

\subsection{Education Mobility} \label{subsec:EM}

In this section we present the results on education mobility across social groups and compare them with results presented in a recent work by \cite{Asher} (Snippet of their results are shown in figure - \ref{fig:AsherResults}). We find that, despite our education mobility estimates depicted in figures - \ref{fig:UME} and \ref{fig:DME}, have coarse confidence bounds in comparison to that in \cite{Asher}, the overall trends are almost the same.\footnote{In fact the similarity of results obtained even with shifted birth cohorts as shown in Figures - \ref{fig:UME_Shift} and \ref{fig:DME_Shift} (in Appendix), with that presented in  \cite{Asher}, further reaffirms the confidence in our results and associated methods used in arriving at them.} We find that the observed mobility patterns for the three common groups SCs, STs, and Muslims, are very closely matching across both the studies. Muslims are observed to be have lowest education mobility prospects particularly in the most recent of cohorts. 

 \begin{table}[!ht]
	\begin{minipage}{.5\columnwidth}
		\begin{Figure}
			\captionsetup{font=scriptsize}
			\begin{center}
				\includegraphics[width=3.5in]{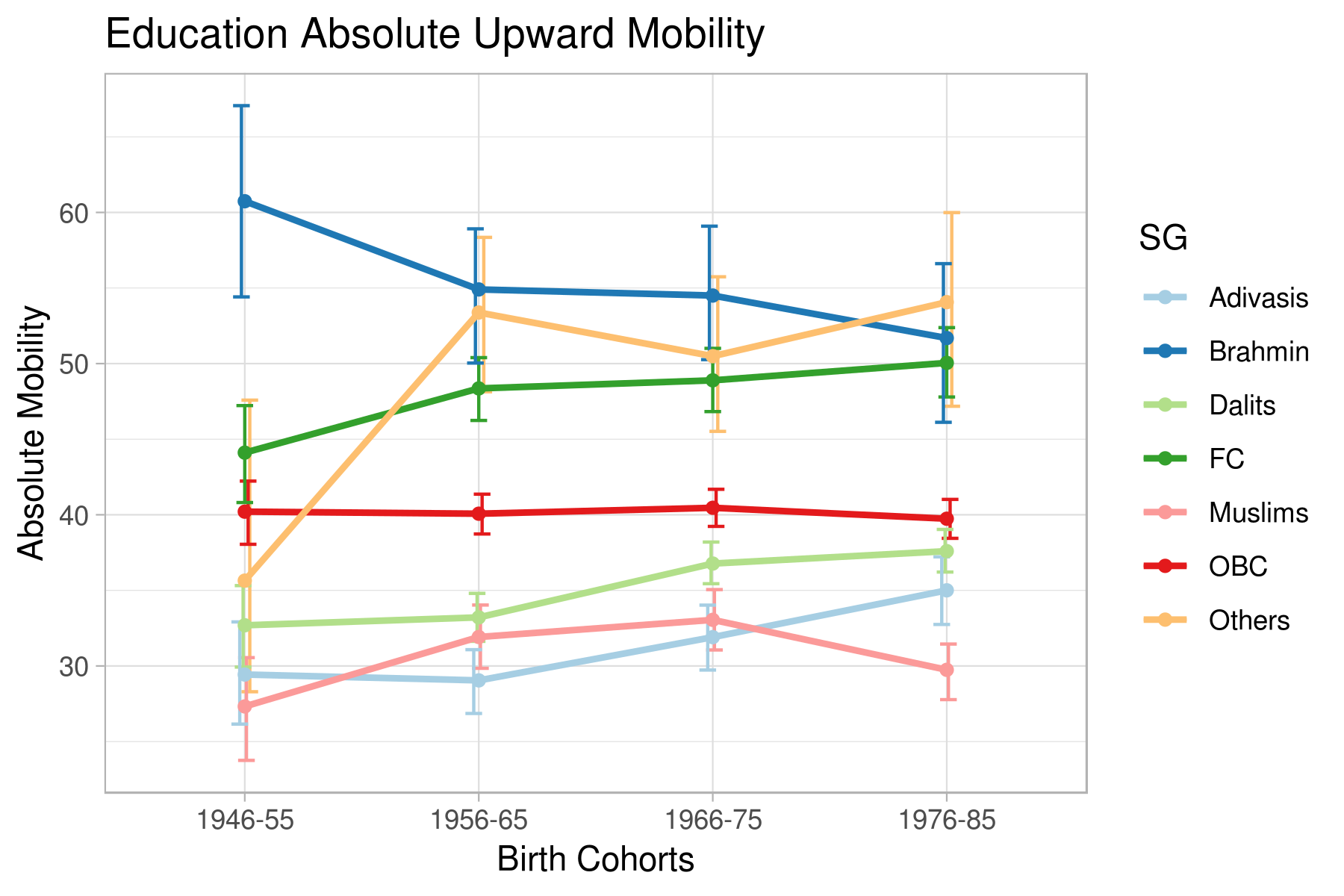}
				\captionof{figure}{Upward Mobility Estimates (Education)} 
				\label{fig:UME}
			\end{center}	
		\end{Figure}
	\end{minipage}
	\begin{minipage}{.5\columnwidth}
		\begin{Figure}
			\captionsetup{font=scriptsize}
			\begin{center}
				\includegraphics[width=3.5in]{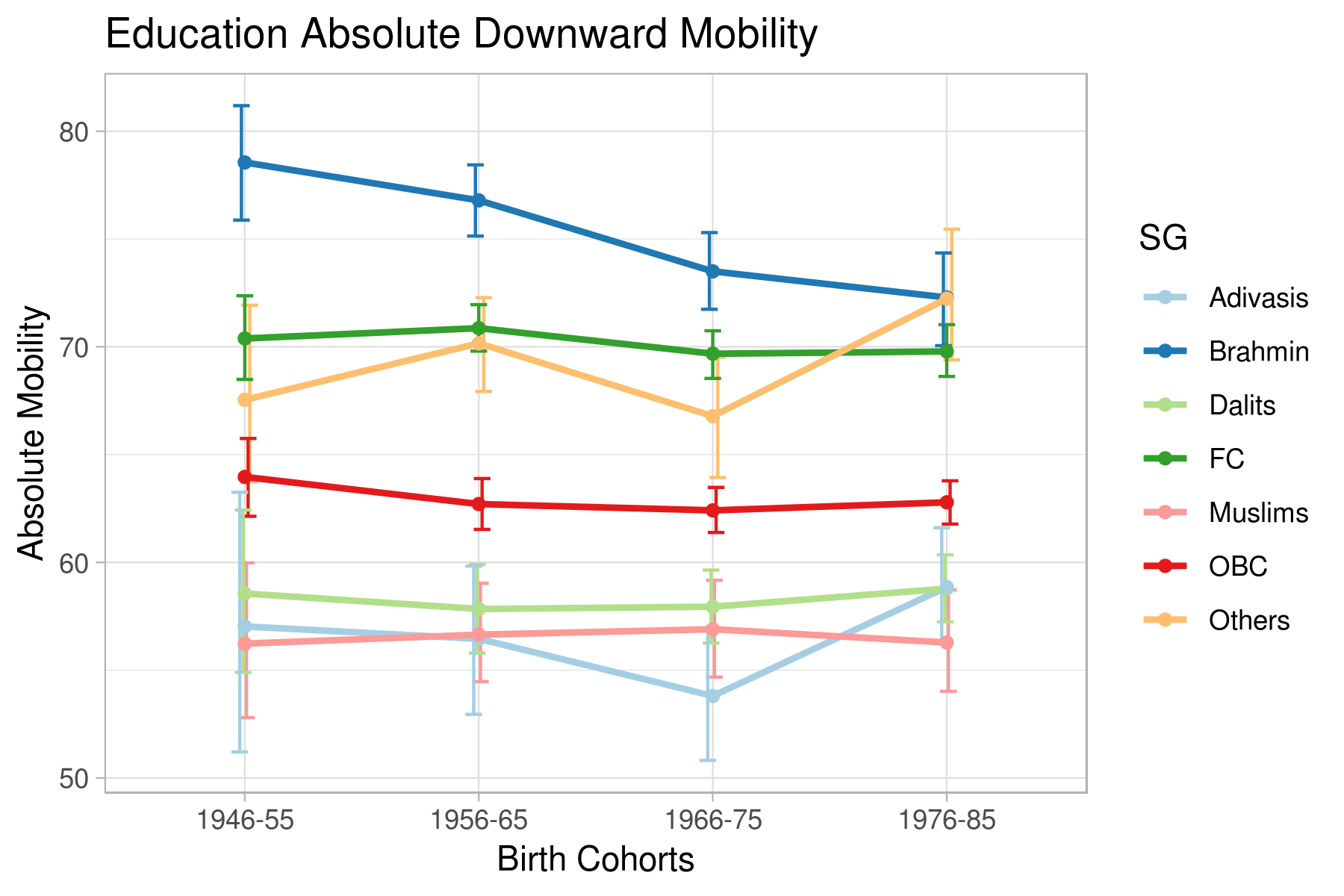}
				\captionof{figure}{Downward Mobility Estimates (Education)} 
				\label{fig:DME}
			\end{center}	
		\end{Figure}
	\end{minipage}
\end{table}

\begin{Figure}
	\captionsetup{font=scriptsize}
	\begin{center}
		\includegraphics[width=6.5in,angle=0]{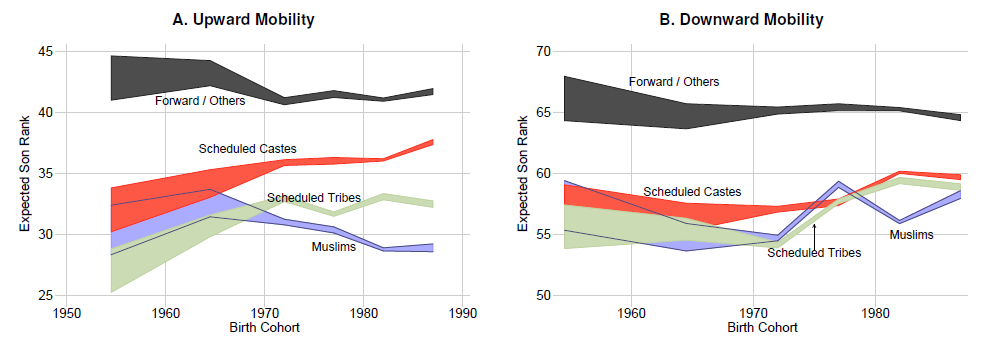}
		\captionof{figure}{A snippet of upward and downward education mobility results presented in \cite{Asher}} 
		\label{fig:AsherResults}
	\end{center}	
\end{Figure}

\subsection{Comparing Occupation and Education Mobility} \label{subsec:contrast}
In the context of our study, socio economic status of occupations indicate the attributes of occupations that can transform education of individual into incomes (\cite{ganzeboom1992standard}). As we have seen earlier in Section - \ref{ComputeSEI}, strength of education in determining the status of occupations, has been on average declining over the years (See Table-\ref{Coeff} or Table-\ref{Coeff_Shift}). In line to some extent with such observation we also find that, despite a rise in education mobility over the years, some social groups do not see an appreciable rise in occupation mobility. This is particularly noticeable when we compare three groups SCs, STs and Muslims. While education mobility seems to indicate that the mobility of SCs and STs is improving over the years in comparison to Muslims, occupation mobility seems to show a slightly contrasting picture. It is observed that despite an increase in upward education mobility, upward occupation mobility for SCs and STs is either decreasing (See Figure - \ref{fig:UMO_Shift}) or stagnant (See Figure - \ref{fig:UMO}). Downward occupation mobility is showing an even worrisome picture, by displaying a decreasing trend for these groups, which is particularly consistent for STs across all robust alternate strategies. These results reaffirm similar findings observed in the works of \cite{motiram2012close} and \cite{iversen2017rags}.\\

How do we reconcile the finding that education mobility differentially affects occupation mobility across social groups? Tables - \ref{tab:STMALL} to \ref{tab:STMST} represent intergenerational occupational mobility observed through father-son transition matrices. We take a finer look at occupational mobility matrices of three groups SCs, STs and Muslims, in order to address our said question. One stark observation indicates that, whether upward or downward transitions, the probability of SCs and STs ending up in occupations codes 6 and 7 (Code-7 here includes 7,8,9 in Table-\ref{tab:1DigSEI}) is considerably high. For Muslims, even though it is comparable, the mobility prospects are also driven by them having a better chance of ending up in occupation code 4 in comparison to the other two groups. As Table-\ref{tab:1DigSEI} indicates, since occupation - 4 is consistently ranked above occupations - 6,7,8 and 9, in terms of both education and income, this may have resulted in better occupational mobility prospects for Muslims over SCs and STs, despite declining education mobility for the former group. It is to be noted that this observation is in no way hinting towards better prospects for Muslims; in fact Figure-\ref{fig:UMO_Shift} shows a declining trend in occupation mobility for the most recent birth cohort, despite an increasing trend observed previously to it (As indicated by Figures - \ref{fig:UMO} and \ref{fig:UMO_Shift}). \\

It seems therefore that, while education mobility measurement is important for observing social mobility, this study suggests that it needs to be sufficiently complemented with other measures to get a near correct picture. A deeper understanding of the nature of occupational transitions across social groups, may point us more towards the actual bottlenecks in realizing upward mobility across such groups. However, for now, we consider this task to be beyond the scope of this paper. 

\section{Discussion}
Importance of intergenerational mobility in its relation to inequality and growth lie in multiple conjectures that have been widely discussed in substantive literature. One dominant proposition associates mobility in a society with its role in palliating the conflicts stemming from growth and redistributive politics, by decreasing the degree of inequality of opportunity for various socio-economic groups. Several studies have contributed to establishing such a proposition over time, a brief summary of which is given in \cite{motiram2012close}. Since inequality of opportunity (measured by its proxy, which is the social mobility) can be analysed along many dimensions (occupation and education being only two among them), trends along one dimension may be completely different in comparison to other dimensions, depending on how the dimensions themselves are related to one other. It is therefore good to complement the observations on mobility along multiple dimensions, and not just one.\\

 In the above study, we observe intergenerational mobility  patterns separately along occupation and education, as the dimensions indicating socio-economic status. We find occupations to be an important dimension, because it closely reflects the socio-economic status of individuals in comparison to either education or income \citep{van2010historical,ganzeboom2003three,ganzeboom1992standard}. As we see through observations in our study, we find sufficient evidence that equality of opportunity in education may not be sufficiently helpful in gaining access into occupations that are socio-economically better positioned. Particularly for SCs and STs, despite an increasing trend observed in education mobility in the recent times, occupational mobility is found to either remain stagnant, or decline, in both cases being much below that of other social groups. There might be structural reasons as to why certain occupations are more accessible to some social groups and difficult for others, which may need further investigation.\\
 
  Results from the study, also hints towards the importance of state's role in  facilitating enabling conditions for individuals across social groups, in translating improvements in education mobility to overall improvement in socio-economic status of individuals. One conspicuous situation we can relate to here, is the state's role in job creation. While the importance of economic growth need not be downplayed, at the same time inadequate creation of jobs by economic-growth cannot also be ignored. The latter may force alternate informal occupations to thrive, and may structurally perpetuate inequalities in access to better occupations, which may be difficult to undo in the future. Inadequate job creation is also a dominant factor in explaining how an unequal society can enable individuals only with connections in gaining the protection of democracy and benefits of economic growth, leaving many in the dim light of poor social mobility prospects \citep{krishna2017broken}. Within the context of growth and distributive justice, characterizing the intergenerational occupational mobility therefore seems to be a crucial precursor to identifying any bottlenecks to equality of opportunity along this dimension. While existing studies, particularly by \cite{Asher}, have observed mobility patterns in education, we find that it is important to complement these studies by also looking at mobility through occupation as the socio-economic status dimension. Nevertheless, it is only a small step taken to further uncover an understanding of the complex causes and consequences of social mobility.

\newpage
\bibliographystyle{apalike}
\bibliography{reflibrary}

\newpage
\section{Appendix} \label{sec:Appendix}
\textbf{Occupational Mobility Matrices}
\begin{table}[!ht]
	\footnotesize
	\caption{State transition matrices capturing occupational mobility across different social groups (computed from IHDS-2).\\
		\footnotesize 	\textit{Note-1: Transition matrices are computed for working sons in the age group 25-55 yrs (or born between 1956-1986), essentially indicating those sons who entered the labour market between 1981-2011. Row names indicate occupation codes of fathers, and column names indicate that of songs.}\\
		\textit{Note-2: n* in all these STM tables indicate the un-weighted number of sons corresponding to fathers belonging to each of the occupational groups (simply indicates the number of such observations in the dataset). However, please note that for computation of transition probabilities, sampling weights are duly considered.}\\
		\scriptsize \textbf{Codes}:- 1 - Professional, Technical and Related Workers; 2 - Administrative Executive and Managerial Workers; 3 - Clerical and Related Workers; 4 - Sales Workers; 5 - Service Workers; 6 - Farmers, agriculture labour, fisherman, hunters and related workers; 7 - Production and Related Workers, Transport Equipment Operators, Labourers; }
	\begin{minipage}{.5\columnwidth}
		\centering
		\begin{tabular}{|r|rrrrrrr|r|}
			\hline
			&  1 & 2 & 3 & 4 & 5 & 6 & 7 & n* \\ 
			\hline
			1& 0.29 & 0.06 & 0.12 & 0.15 & 0.04 & 0.13 & 0.21 & 759\\ 
			2& 0.10 & 0.31 & 0.07 & 0.16 & 0.04 & 0.07 & 0.25 & 370\\ 
			3& 0.11 & 0.07 & 0.29 & 0.13 & 0.04 & 0.11 & 0.25 & 921\\ 
			4& 0.05 & 0.07 & 0.04 & 0.48 & 0.03 & 0.06 & 0.26 & 1305\\ 
			5& 0.06 & 0.04 & 0.05 & 0.08 & 0.33 & 0.16 & 0.28 & 1145\\ 
			6 & 0.03 & 0.02 & 0.03 & 0.05 & 0.03 & 0.58 & 0.26 & 15324\\ 
			7 & 0.03 & 0.04 & 0.05 & 0.08 & 0.04 & 0.10 & 0.67 & 5142\\ 
			\hline
		\end{tabular}
		\caption{STM ALL}
		\label{tab:STMALL}
	\end{minipage}
	\begin{minipage}{.5\columnwidth}
		\centering
		\begin{tabular}{|r|rrrrrrr|r|}
			\hline
			&  1 & 2 & 3 & 4 & 5 & 6 & 7 & n* \\ 
			\hline
			1 & 0.36 & 0.08 & 0.14 & 0.13 & 0.03 & 0.10 & 0.15 & 357\\ 
			2& 0.16 & 0.39 & 0.09 & 0.14 & 0.02 & 0.02 & 0.17 & 124\\ 
			3 & 0.17 & 0.07 & 0.32 & 0.10 & 0.03 & 0.08 & 0.23 & 401\\ 
			4 & 0.06 & 0.10 & 0.08 & 0.48 & 0.02 & 0.05 & 0.21 & 449\\ 
			5 & 0.12 & 0.06 & 0.10 & 0.12 & 0.22 & 0.16 & 0.22 & 295\\ 
			6 & 0.05 & 0.04 & 0.07 & 0.07 & 0.04 & 0.55 & 0.19 & 3061\\ 
			7 & 0.04 & 0.06 & 0.08 & 0.12 & 0.05 & 0.07 & 0.58 & 759\\ 
			\hline
		\end{tabular}
		\caption{STM OTH}
		\label{tab:STMOTH}
	\end{minipage}
\end{table}

\begin{table}[!ht]
	\footnotesize
	\begin{minipage}{.5\columnwidth}
		\centering
		\begin{tabular}{|r|rrrrrrr|r|}
			\hline
			&  1 & 2 & 3 & 4 & 5 & 6 & 7 & n* \\ 
			\hline
			1  & 0.22 & 0.03 & 0.11 & 0.18 & 0.03 & 0.19 & 0.24 & 197\\ 
			2 & 0.06 & 0.30 & 0.08 & 0.14 & 0.07 & 0.10 & 0.25 & 94\\ 
			3  & 0.05 & 0.05 & 0.24 & 0.19 & 0.04 & 0.17 & 0.27 & 222\\ 
			4  & 0.05 & 0.05 & 0.03 & 0.48 & 0.03 & 0.08 & 0.27 & 401\\ 
			5  & 0.05 & 0.07 & 0.03 & 0.07 & 0.32 & 0.18 & 0.28 & 373\\ 
			6 & 0.02 & 0.02 & 0.03 & 0.05 & 0.03 & 0.61 & 0.25 & 5600\\ 
			7 & 0.02 & 0.04 & 0.06 & 0.08 & 0.04 & 0.08 & 0.68 & 1641\\ 
			\hline
		\end{tabular}
		\caption{STM OBC}
		\label{tab:STMOBC}
	\end{minipage}
	\begin{minipage}{.5\columnwidth}
		\centering
		\begin{tabular}{|r|rrrrrrr|r|}
			\hline
			&  1 & 2 & 3 & 4 & 5 & 6 & 7 & n* \\ 
			\hline
			1 & 0.18 & 0.05 & 0.07 & 0.26 & 0.03 & 0.12 & 0.29 & 81\\ 
			2 & 0.03 & 0.34 & 0.02 & 0.28 & 0.03 & 0.04 & 0.25 & 75\\ 
			3 & 0.11 & 0.08 & 0.21 & 0.16 & 0.02 & 0.03 & 0.38 & 96\\ 
			4  & 0.04 & 0.05 & 0.02 & 0.52 & 0.03 & 0.05 & 0.30 & 302\\ 
			5 & 0.04 & 0.04 & 0.01 & 0.12 & 0.27 & 0.11 & 0.42 & 95\\ 
			6  & 0.02 & 0.01 & 0.02 & 0.08 & 0.01 & 0.49 & 0.36 & 1322\\ 
			7  & 0.02 & 0.06 & 0.02 & 0.11 & 0.03 & 0.06 & 0.70 & 890\\ 
			\hline
		\end{tabular}
		\caption{STM MUS}
		\label{tab:STMMUS}
	\end{minipage}
\end{table}

\begin{table}[!ht]
	\footnotesize
	\begin{minipage}{.5\columnwidth}
		\centering
		\begin{tabular}{|r|rrrrrrr|r|}
			\hline
			&  1 & 2 & 3 & 4 & 5 & 6 & 7 & n* \\ 
			\hline
			1& 0.22 & 0.04 & 0.11 & 0.10 & 0.08 & 0.09 & 0.36 & 99\\ 
			2& 0.11 & 0.16 & 0.11 & 0.09 & 0.04 & 0.09 & 0.40 & 66\\ 
			3& 0.07 & 0.06 & 0.38 & 0.08 & 0.08 & 0.10 & 0.22 & 166\\ 
			4& 0.02 & 0.08 & 0.03 & 0.46 & 0.04 & 0.07 & 0.31 & 128\\ 
			5 & 0.02 & 0.01 & 0.06 & 0.03 & 0.44 & 0.14 & 0.29 & 335\\ 
			6 & 0.02 & 0.01 & 0.03 & 0.03 & 0.04 & 0.54 & 0.33 & 3439\\ 
			7 & 0.02 & 0.02 & 0.03 & 0.05 & 0.04 & 0.13 & 0.70 & 1527\\ 
			\hline
		\end{tabular}
		\caption{STM SC}
		\label{tab:STMSC}
	\end{minipage}
	\begin{minipage}{.5\columnwidth}
		\centering
		\begin{tabular}{|r|rrrrrrr|r|}
			\hline
			&  1 & 2 & 3 & 4 & 5 & 6 & 7 & n* \\ 
			\hline
			1  & 0.35 & 0.09 & 0.21 & 0.05 & 0.06 & 0.18 & 0.06 & 25\\ 
			2 & 0.08 & 0.18 & 0.00 & 0.00 & 0.06 & 0.42 & 0.26 & 11\\ 
			3  & 0.05 & 0.09 & 0.23 & 0.00 & 0.13 & 0.25 & 0.26 & 36\\ 
			4  & 0.06 & 0.09 & 0.00 & 0.43 & 0.18 & 0.05 & 0.20 & 25\\ 
			5  & 0.06 & 0.02 & 0.06 & 0.07 & 0.29 & 0.24 & 0.26 & 47\\ 
			6  & 0.02 & 0.01 & 0.02 & 0.02 & 0.04 & 0.68 & 0.21 & 1902\\ 
			7  & 0.03 & 0.02 & 0.03 & 0.03 & 0.03 & 0.23 & 0.63 & 325\\ 
			\hline
		\end{tabular}
		\caption{STM ST}
		\label{tab:STMST}
	\end{minipage}
\end{table}

\newpage
\textbf{Results by Shifting Cohorts}
\begin{table}[!ht]
	\begin{minipage}{.5\columnwidth}
		\begin{Figure}
			\captionsetup{font=scriptsize}
			\begin{center}
				\includegraphics[width=3.5in]{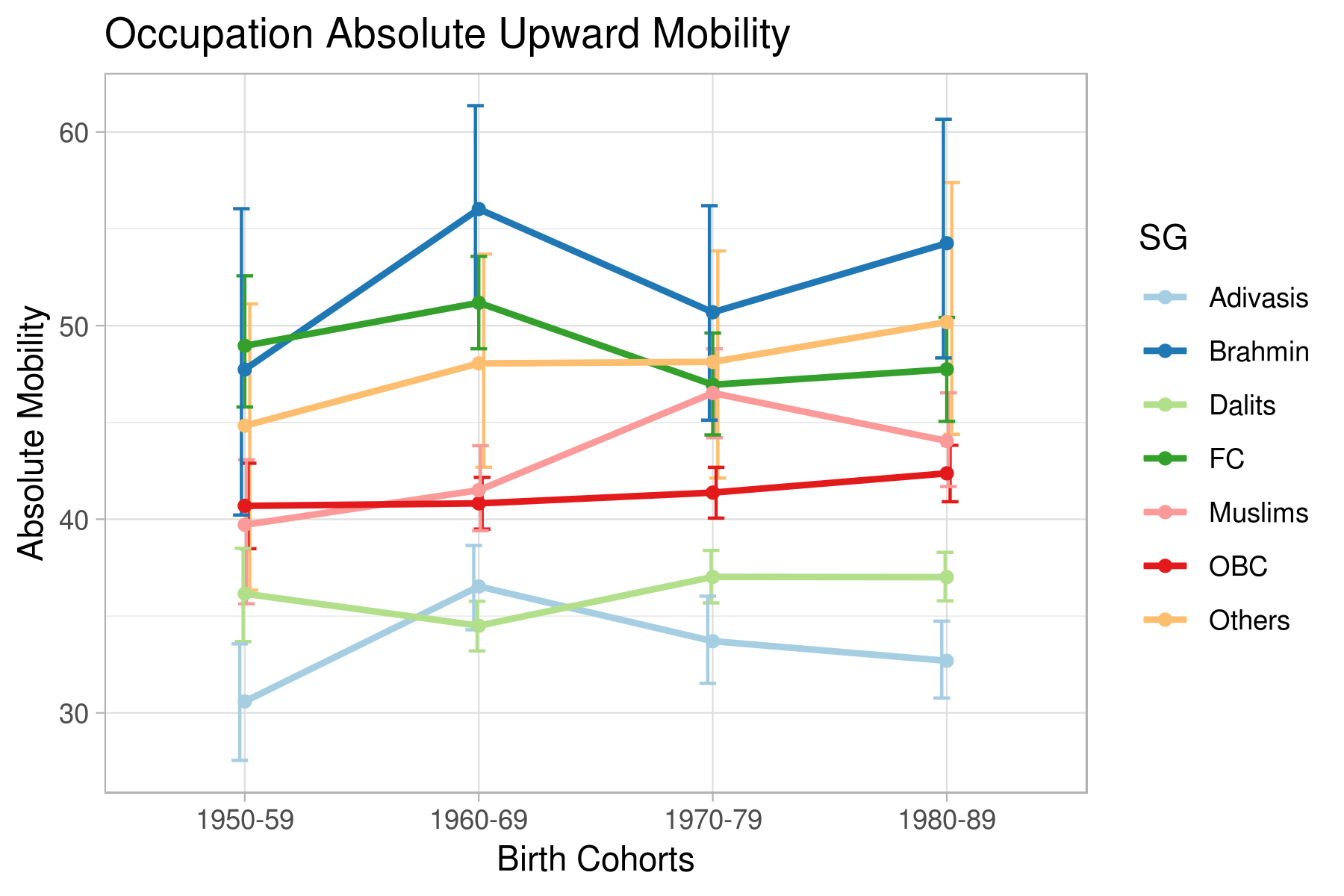}
				\captionof{figure}{Upward Mobility Estimates (Occupation) - Shifted Cohorts} 
				\label{fig:UMO_Shift}
			\end{center}	
		\end{Figure}
	\end{minipage}
	\begin{minipage}{.5\columnwidth}
		\begin{Figure}
			\captionsetup{font=scriptsize}
			\begin{center}
				\includegraphics[width=3.5in]{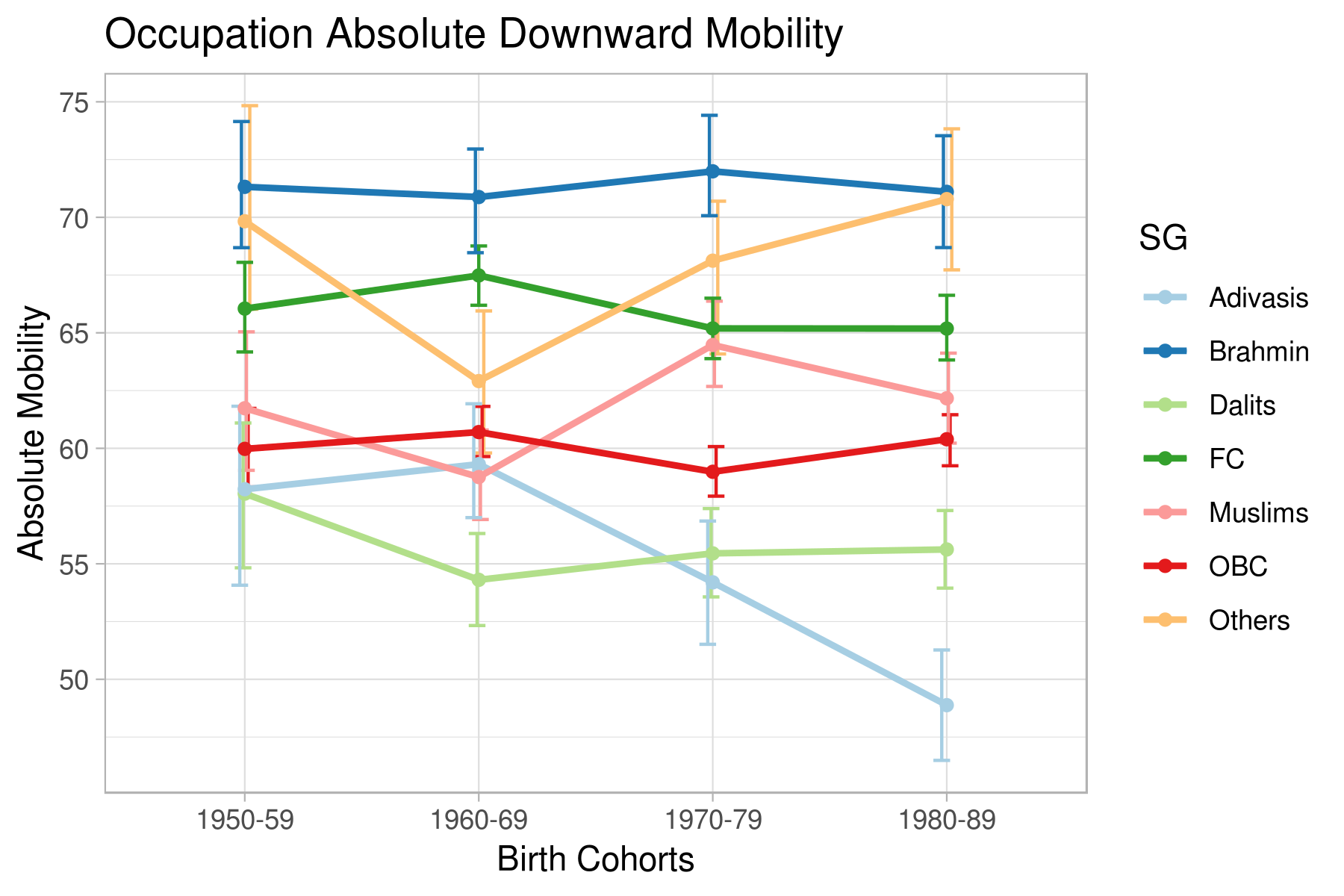}
				\captionof{figure}{Downward Mobility Estimates (Occupation) - Shifted Cohorts} 
				\label{fig:DMO_Shift}
			\end{center}	
		\end{Figure}
	\end{minipage}
\end{table}

\begin{table}[!ht]
	\begin{minipage}{.5\columnwidth}
		\begin{Figure}
			\captionsetup{font=scriptsize}
			\begin{center}
				\includegraphics[width=3.5in]{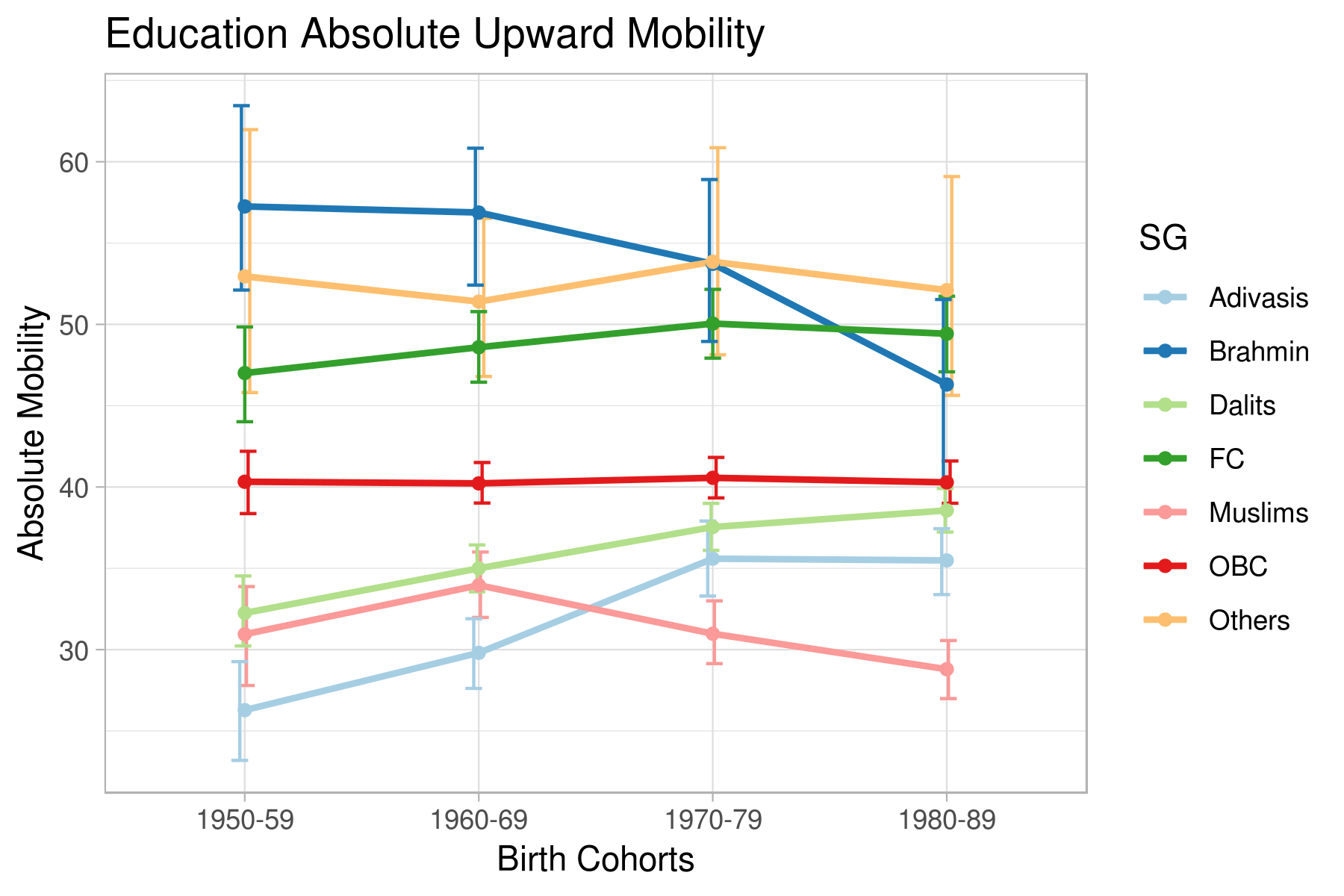}
				\captionof{figure}{Upward Mobility Estimates (Education) - Shifted Cohorts} 
				\label{fig:UME_Shift}
			\end{center}	
		\end{Figure}
	\end{minipage}
	\begin{minipage}{.5\columnwidth}
		\begin{Figure}
			\captionsetup{font=scriptsize}
			\begin{center}
				\includegraphics[width=3.5in]{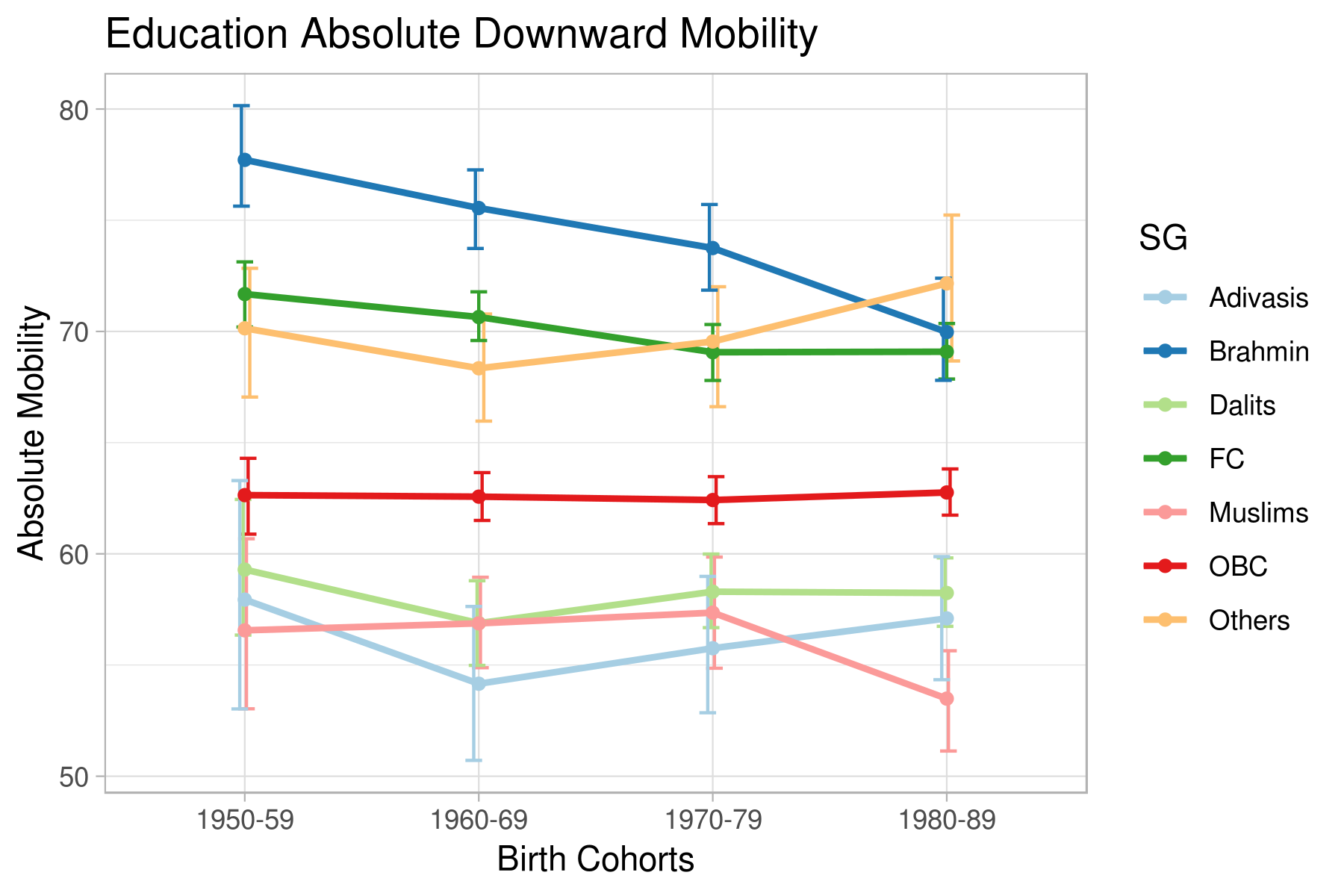}
				\captionof{figure}{Downward Mobility Estimates (Education) - Shifted Cohorts} 
				\label{fig:DME_Shift}
			\end{center}	
		\end{Figure}
	\end{minipage}
\end{table}

\begin{table}[!ht]
	\small
	\begin{minipage}{.5\columnwidth}
	\centering
	\begin{tabular}{|p{3cm}|p{2cm}|p{2cm}|}
		\hline
		\textbf{Birth-Cohort} & $\beta_{43}$ & $\beta_{32}$ \\ 
		\hline
		1930-39 & 0.79 & 0.35 \\
		\hline
		1940-49 &0.79&0.37  \\ 
		\hline
		1950-59 &0.80  & 0.35 \\
		\hline
		1960-69 &  0.80&0.26  \\ 
		\hline
		1970-79&0.83&0.24\\
		\hline
		1980-89 &0.88&0.19\\
		\hline
	\end{tabular}
	\caption{\label{Coeff_Shift} Strength of Income and Education in determining SEI score for shifted cohorts\\ \scriptsize (Refer Step-3 in Table-\ref{SAMAlgo})}
		\end{minipage}
	\begin{minipage}{.5\columnwidth}
		\begin{tabular}{|r|l|c|c|c|}
			\hline
			& Cohort & Pearson & Spearman & Kendall \\  
			\hline
			1 & 1930-39 & 0.00 & 0.00 & 0.00 \\ 
			2 & 1940-49 & 0.95 & 0.94 & 0.80 \\ 
			3 & 1950-59 & 0.96 & 0.95 & 0.84 \\ 
			4 & 1960-69 & 0.91 & 0.90 & 0.76 \\ 
			5 & 1970-79 & 0.81 & 0.81 & 0.62 \\ 
			6 & 1980-89 & 0.69 & 0.70 & 0.54 \\ 
			\hline
		\end{tabular}
		\caption{\label{Corr_Shift} \scriptsize Correlation of Socio-Economic Status Rankings of occupations across new birth cohorts  [Corr($S_t,S_{t-1}$)] \\ 
			\scriptsize  Note:  $S_t$ is the vector of socio economic status ranking of occupations corresponding to a birth cohort and $S_{t-1}$ indicates its previous cohort.}
	\end{minipage}
\end{table}

\newpage
\begin{table}[ht]
	\centering
	\scriptsize
	\begin{adjustbox}{width=1\textwidth}
		\begin{tabular}{|p{0.5cm}|p{0.5cm}|p{15cm}|p{0.5cm}|p{0.5cm}|p{0.5cm}|}
			\hline
			(1) & (2) & (3) & (4) & (5) & (6) \\ 
			\hline
			0 & 2 & Architects, Engineers, Technologists and Surveyors & 75 & 80 & 75 \\ 
			0 & 3 & Engineering Technicians & 65 & 74 & 69 \\ 
			0 & 7 & Physicians and Surgeons (Allopathic Dental and Veterinary Surgeons) & 63 & 73 & 73 \\ 
			0 & 8 & Nursing and other Medical and Health Technicians & 51 & 56 & 64 \\ 
			0 & 9 & Scientific, Medical and Technical Persons, Other & 52 & 63 & 58 \\ 
			1 & 12 & Accountants, Auditors and Related Workers & 68 & 75 & 76 \\ 
			1 & 14 & Jurists & 70 & 76 & 80 \\ 
			1 & 15 & Teachers & 56 & 59 & 72 \\ 
			1 & 16 & Poets, Authors, Journalists and Related Workers & 63 & 71 & 77 \\ 
			1 & 17 & Sculptors, Painters, Photographers and Related Creative Artists & 44 & 53 & 47 \\ 
			1 & 18 & Composers and Performing Artists & 29 & 30 & 27 \\ 
			1 & 19 & Professional Workers, n.e.c. & 30 & 17 & 55 \\ 
			2 & 21 & Administrative and Executive Officials Government and Local Bodies & 71 & 78 & 78 \\ 
			2 & 22 & Working Proprietors, Directors and Managers, Wholesale and Retail Trade & 51 & 67 & 50 \\ 
			2 & 23 & Directors and Managers, Financial Institutions & 71 & 77 & 79 \\ 
			2 & 24 & Working Proprietors, Directors and Managers Mining, Construction, Manufacturing and Related Concerns & 56 & 72 & 57 \\ 
			2 & 25 & Working Proprietors, Directors, Managers and Related Executives, Transport, Storage and Communication & 53 & 70 & 46 \\ 
			2 & 26 & Working Proprietors, Directors and Managers, Other Service & 48 & 58 & 52 \\ 
			2 & 29 & Administrative, Executive and Managerial Workers, n.e.c. & 71 & 79 & 74 \\ 
			3 & 30 & Clerical and Related Workers & 60 & 69 & 70 \\ 
			3 & 31 & Village Officials & 41 & 40 & 61 \\ 
			3 & 32 & Stenographers, Typists and Card and Tape Punching Operators & 60 & 68 & 71 \\ 
			3 & 33 & Book-keepers, Cashiers and Related Workers & 53 & 55 & 68 \\ 
			3 & 34 & Computing Machine Operators & 56 & 66 & 66 \\ 
			3 & 35 & Clerical and Related Workers, n.e.c. & 46 & 52 & 60 \\ 
			3 & 36 & Transport and Communication Supervisors & 54 & 60 & 67 \\ 
			3 & 37 & Transport Conductors and Guards & 43 & 48 & 54 \\ 
			3 & 38 & Mail Distributors and Related Workers & 42 & 46 & 56 \\ 
			3 & 39 & Telephone and Telegraph Operators & 54 & 61 & 65 \\ 
			4 & 40 & Merchants and Shopkeepers, Wholesale and Retail Trade & 37 & 44 & 39 \\ 
			4 & 41 & Manufacturers, Agents & 49 & 54 & 62 \\ 
			4 & 42 & Technical Salesmen and Commercial Travellers & 53 & 62 & 63 \\ 
			4 & 43 & Salesmen, Shop Assistants and Related Workers & 26 & 21 & 30 \\ 
			4 & 44 & Insurance, Real Estate, Securities and Business Service Salesmen and Auctioneers & 52 & 65 & 59 \\ 
			4 & 45 & Money Lenders and Pawn Brokers & 47 & 57 & 51 \\ 
			4 & 49 & Sales Workers, n.e.c. & 25 & 22 & 28 \\ 
			5 & 50 & Hotel and Restaurant Keepers & 32 & 36 & 32 \\ 
			5 & 52 & Cooks, Waiters, Bartenders and Related Worker (Domestic and Institutional) & 16 & 5 & 24 \\ 
			5 & 53 & Maids and Other House Keeping Service Workers n.e.c. & 16 & 9 & 13 \\ 
			5 & 54 & Building Caretakers, Sweepers, Cleaners and Related Workers & 25 & 29 & 12 \\ 
			5 & 55 & Launderers, Dry-cleaners and Pressers & 18 & 15 & 11 \\ 
			5 & 56 & Hair Dressers, Barbers, Beauticians and Related Workers & 17 & 12 & 17 \\ 
			5 & 57 & Protective Service Workers & 40 & 42 & 49 \\ 
			5 & 59 & Service Workers, n.e.c. & 27 & 23 & 31 \\ 
			6 & 60 & Farm Plantation, Dairy and Other Managers and Supervisors & 38 & 45 & 36 \\ 
			6 & 61 & Cultivators & 25 & 27 & 19 \\ 
			6 & 62 & Farmers other than Cultivators & 29 & 34 & 21 \\ 
			6 & 63 & Agricultural Labourers & 3 & 1 & 1 \\ 
			6 & 64 & Plantation Labourers and Related Workers & 9 & 2 & 3 \\ 
			6 & 65 & Other Farm Workers & 15 & 10 & 9 \\ 
			6 & 66 & Forestry Workers & 18 & 13 & 18 \\ 
			6 & 68 & Fishermen and Related Workers & 21 & 19 & 7 \\ 
			7 & 71 & Miners, Quarrymen, Well Drillers and Related Workers & 22 & 20 & 16 \\ 
			7 & 72 & Metal Processors & 32 & 33 & 33 \\ 
			7 & 73 & Wood Preparation Workers and Paper Makers & 26 & 28 & 20 \\ 
			7 & 74 & Chemical Processors and Related Workers & 50 & 64 & 44 \\ 
			7 & 75 & Spinners, Weavers, Knitters, Dyers and Related Workers & 23 & 18 & 22 \\ 
			7 & 77 & Food and Beverage Processors & 25 & 25 & 25 \\ 
			7 & 78 & Tobacco Preparers and Tobacco Product Makers & 12 & 4 & 5 \\ 
			7 & 79 & Tailors, Dress Makers, Sewers, Upholsterers and Related Workers & 29 & 32 & 29 \\ 
			8 & 80 & Shoe makers and Leather Goods Makers & 18 & 14 & 14 \\ 
			8 & 81 & Carpenters, Cabinet and Related Wood Workers & 25 & 26 & 23 \\ 
			8 & 82 & Stone Cutters and Carvers & 15 & 11 & 4 \\ 
			8 & 83 & Blacksmiths, Tool Makers and Machine Tool Operators & 32 & 35 & 34 \\ 
			8 & 84 & Machinery Fitters, Machine Assemblers and Precision Instrument Makers (except Electrical) & 40 & 47 & 42 \\ 
			8 & 85 & Electrical Fitters and Related Electrical and Electronic Workers & 42 & 50 & 45 \\ 
			8 & 86 & Broadcasting Station and Sound Equipment Operators and Cinema Projectionists & 40 & 43 & 53 \\ 
			8 & 87 & Plumbers, Welders, Sheet Metal and Structural Metal Preparers and Erectors & 34 & 38 & 35 \\ 
			8 & 88 & Jewellery and Precious Metal Workers and Metal Engravers (Except Printing) & 40 & 51 & 40 \\ 
			8 & 89 & Glass Formers, Potters and Related Workers & 12 & 6 & 2 \\ 
			9 & 90 & Rubber and Plastic Product Makers & 36 & 37 & 43 \\ 
			9 & 91 & Paper and Paper Board Products Makers & 36 & 41 & 38 \\ 
			9 & 92 & Printing and Related Workers & 38 & 39 & 48 \\ 
			9 & 93 & Painters & 28 & 24 & 37 \\ 
			9 & 94 & Production and Related Workers, n.e.c. & 14 & 7 & 10 \\ 
			9 & 95 & Bricklayers and Other Constructions Workers & 15 & 8 & 8 \\ 
			9 & 96 & Stationery Engines and Related Equipment Operators, Oilers and Greasers & 39 & 49 & 41 \\ 
			9 & 97 & Material Handling and Related Equipment Operators, Loaders and Unloaders & 20 & 16 & 15 \\ 
			9 & 98 & Transport Equipment Operators & 28 & 31 & 26 \\ 
			9 & 99 & Labourers, n.e.c. & 11 & 3 & 6 \\ 
			\hline
		\end{tabular}
	\end{adjustbox}
	\caption{\small Detailed Summary of average SEI scores, Income and Education Ranks, across occupations\\
		\scriptsize Column Index: (1) - One Digit Code, (2) - Two Digit Code,(3) - Description, (4) - Average SEI score across cohorts (1-80),(5) - Income Rank (1-80), (6) - Education Rank (1-80); Ranks are in ascending order of income/education levels; Note that these results are for our original cohort scheme in Table-\ref{Cohorts}}
	\label{tab:FullSummary}
\end{table}

\end{document}